%% file: ms.tex
\documentclass[times, twocolumn]{aastex63}
\usepackage{graphicx, graphics}
\usepackage{multirow}
\usepackage{CJK}
\newcommand\numberthis{\addtocounter{equation}{1}\tag{\theequation}}
\usepackage{chngcntr}
\usepackage{bbm}
\usepackage[nodayofweek]{datetime}
\newdateformat{monthyearday}{%
  \THEYEAR\ \monthname[\THEMONTH] \twodigit{\THEDAY}}
  
\usepackage{amsmath, amssymb, bm}

\newcommand{\Qr}{$\mathcal{Q}_\phi$}
\newcommand{\Ur}{$\mathcal{U}_\phi$}

\received{2018 December 11}
\revised{2018 July 18}
\accepted{2018 July 18}
\acceptjournall{\textit{The Astrophysical Journal}}
\notetoeditor{text}
\shorttitle{HD~191089 disk in Hubble + Gemini}
\shortauthors{Ren et al.}

\begin{document}
\begin{CJK*}{UTF8}{gbsn}
\title{An Exo--Kuiper Belt with An Extended Halo around HD~191089 in Scattered Light}
\author[0000-0003-1698-9696]{Bin Ren (任彬)}\email{ren@jhu.edu}
\affiliation{Department of Physics and Astronomy, The Johns Hopkins University, 3701 San Martin Drive, Baltimore, MD 21218, USA}
\affiliation{Department of Applied Mathematics and Statistics, The Johns Hopkins University, 3400 North Charles Street, Baltimore, MD 21218, USA}

\author[0000-0002-9173-0740]{\'Elodie Choquet} 
\altaffiliation{NASA Hubble Fellow}
\affiliation{Aix Marseille Univ, CNRS, CNES, LAM, Marseille, France}
\affiliation{Department of Astronomy, California Institute of Technology, 1200 East California Boulevard, Pasadena, CA 91125, USA}
\affiliation{NASA Jet Propulsion Laboratory, California Institute of Technology, 4800 Oak Grove Drive, Pasadena, CA 91109, USA}

\author[0000-0002-3191-8151]{Marshall D. Perrin} 
\affiliation{Space Telescope Science Institute (STScI), 3700 San Martin Drive, Baltimore, MD 21218, USA}

\author[0000-0002-5092-6464]{Gaspard Duch\^ene} 
\affiliation{Astronomy Department, University of California, Berkeley, CA 94720, USA}
\affiliation{Universit\'e Grenoble-Alpes, CNRS Institut de Plan\'etologie et d'Astrophysique (IPAG), F-38000 Grenoble, France}

\author[0000-0002-1783-8817]{John H. Debes} 
\author{Laurent Pueyo} 
\affiliation{Space Telescope Science Institute (STScI), 3700 San Martin Drive, Baltimore, MD 21218, USA}

\author[0000-0002-7670-670X]{Malena Rice}
\affiliation{Department of Astronomy, Yale University, New Haven, CT 06511, USA}

\author[0000-0002-8382-0447]{Christine Chen} 
\affiliation{Space Telescope Science Institute (STScI), 3700 San Martin Drive, Baltimore, MD 21218, USA}
\affiliation{Department of Physics and Astronomy, The Johns Hopkins University, 3701 San Martin Drive, Baltimore, MD 21218, USA}

\author[0000-0002-4511-5966]{Glenn Schneider}
\affiliation{Steward Observatory, The University of Arizona, 933 North Cherry Avenue, Tucson, AZ 85721, USA}

\author[0000-0002-0792-3719]{Thomas M. Esposito}
\affiliation{Astronomy Department, University of California, Berkeley, CA 94720, USA}
 
\author[0000-0003-4845-7483]{Charles A. Poteet}  
\affiliation{Space Telescope Science Institute (STScI), 3700 San Martin Drive, Baltimore, MD 21218, USA}

\author[0000-0003-0774-6502]{Jason J. Wang}
\altaffiliation{51 Pegasi b Fellow}
\affiliation{Department of Astronomy, California Institute of Technology, 1200 East California Boulevard, Pasadena, CA 91125, USA}

\input{AlphabetAuthors.txt}

\author[0000-0002-5222-5717]{Colin Norman}
\affiliation{Department of Physics and Astronomy, The Johns Hopkins University, 3701 San Martin Drive, Baltimore, MD 21218, USA}
\affiliation{Space Telescope Science Institute (STScI), 3700 San Martin Drive, Baltimore, MD 21218, USA}

\begin{abstract}
We have obtained {\it Hubble Space Telescope} ({\it HST}) STIS and NICMOS, and {\it Gemini}/GPI scattered light images of the HD~191089 debris disk. We identify two spatial components: a ring resembling Kuiper Belt in radial extent (FWHM: ${\sim}25$ au, centered at ${\sim}46$ au), and a halo extending to ${\sim}640$~au. We find that the halo is significantly bluer than the ring, consistent with the scenario that the ring serves as the ``birth ring'' for the smaller dust in the halo. We measure the scattering phase functions in the $30^\circ$--$150^\circ$ scattering angle range and find the halo dust is both more forward- and backward-scattering than the ring dust.  We measure a surface density power law index of ${-0.68\pm0.04}$ for the halo, which indicates the slow-down of the radial outward motion of the dust. {Using radiative transfer modeling, we attempt to simultaneously reproduce the (visible) total and (near-infrared) polarized intensity images of the birth ring. Our modeling leads to mutually inconsistent results, indicating that more complex models, such as the inclusion of more realistic aggregate particles, are needed.}
\end{abstract}


\section{Introduction}

Debris disks, the extrasolar analogs of the asteroid belt and Kuiper Belt, have been detected around ${\sim}20\%$ of the nearest stars (A-type stars: \citealp{thureau14}; FGK stars: \citealp{eiroa13, montesinos16, sibthorpe18}). They are expected to be the results from the grinding down of larger dust \citep{wyatt08}, {however the diversity in observables such as morphology and surface brightness suggests that they are shaped by a variety of mechanisms \citep[e.g.,][]{artymowicz97, stark14, lee16}}. Imaging studies of debris disks in scattered light use not only space-based instruments (e.g., STIS: \citealp{schneider09, schneider14, schneider16, schneider18, konishi16}; NICMOS: \citealp{soummer14, choquet16, choquet17, choquet18}) that offer the best telescope stability, but also extreme adaptive optics--equipped ground-based instruments (e.g., GPI: \citealp{perrin15, kalas15, hung15, draper16, millarblanchaer15, millarblanchaer16, esposito18}; SPHERE: \citealp{boccaletti15, lagrange16, wahhaj16, feldt17, milli17, matthews17, engler17, sissa18, olofsson18, milli19}) that provide the best angular resolution and probe closer-in regions of the disks.

Multi-wavelength studies can provide complementary insights in understanding circumstellar disks, since different wavelengths probe distinct regions and parameter space for a disk \citep{ertel12, siciliaaguilar16}. In this paper, we focus on {using scattered light observations to understand one of these systems}. In previous studies, the combination of both  space- and ground-based instruments has been implemented to study both protoplanetary \citep[e.g., PDS 66: ][]{wolff16} and debris disks (e.g., 49 Ceti: \citealp{choquet17}, HD 35841: \citealp{esposito18}), and those observations are interpreted using radiative transfer codes \citep[e.g.,][]{augereau06, milli15, wolff17, esposito18}. We perform such an analysis for the debris disk surrounding HD~191089 to study its specific properties via measurement and radiative transfer modeling in this paper.

We list the properties of the system in Table~\ref{tab:star}: HD~191089 is an F5V star with $T_{\text{eff}} = 6450$ K located at $50.14\pm0.11$~pc \citep[{\it Gaia} DR2:][]{gaia18}. \citet{moor06} identified it as a member of the Beta Pictoris moving group with space velocities compatible with the kinematics of the group, and \citet{shkolnik17} {estimated the age of this group to be $22\pm6$~Myr based on the consensus of the group members}.

\begin{deluxetable}{lrc}[htb!]
\tablecaption{System properties  \label{tab:star}}
\tablehead{\colhead{Properties}	& \colhead{HD 191089} & \colhead{Reference}}
\startdata
Distance (pc)	&$50.14\pm0.11$	&1	\\
RA (J2000) 		&  20 09 05.215	&1\\
Dec (J2000)		& -26 13 26.520	&1\\
Spectral Type	&F5V 			&2, 3\\
$M_\star$ ($M_\odot$) & $1.4\pm0.1$		& 4\\
$T_{\rm eff}$ (K)	& $6450$ 			&1\\
$V$ (mag)			& $7.18$			&5\\
$J$ (mag)			& $6.321$  		&6\\
$H$ (mag)			& $6.091$	 		&6\\
Association 	&$\beta$ Pic Moving Group	&7\\
Age (Myr)		&$22\pm6$		&8\\
$L_{\rm dust}/L_\star$&$(14.2\pm0.5)\times10^{-4}$	&9\\
Proper Motion (RA) & $40.17\pm0.07$ mas yr$^{-1}$ & 1\\
Proper Motion (Dec) & $-67.38\pm0.05$ mas yr$^{-1}$ & 1\\
Radial Velocity &	$-5.4\pm0.4$ km s$^{-1}$ & 1 \\
\enddata
\tablerefs{1: \citet{gaia18}; 2: \citet{houk82}; 3: \citet{hales17}; 4: \citet{chandler16}; 5: \citet{hog00}; 6: \citet{cutri03}; 7: \citet{moor06}; 8: \citet{shkolnik17}; 9: \citet{holland17}.}
\end{deluxetable} 
 
Before a resolved scattered light image was reported, HD~191089 was first identified by \citet{mannings98} as a debris disk candidate based on {\it IRAS} infrared excess. \citet{chen14} suggested a two-temperature model to explain the Spectral Energy Distribution (SED) of the system, while \citet{kennedy14} argued for one temperature. The latest SED analysis including {\it CSO}, {\it Herschel}, and {\it JCMT} photometry up to 850 $\mu$m seems to confirm the latter hypothesis with a best fit obtained using a single-component disk: assuming the dust behaves as a blackbody, SED analysis suggests a dust mass ${\sim}0.037 M_\earth$, a temperature of $89$~K, and a radius ${\sim}17$~au \citep{holland17}. However, as is commonly seen for many disks, the radius derived assuming blackbody dust is several times smaller than the radius observed in resolved images (e.g., SEDs: \citealp{mittal15}; images: \citealp{hughes18}, and references therein). This indicates that the dust grains are not simple black bodies, {and the smallest dust are not efficient emitters at long wavelengths}. 

The HD~191089 disk was first resolved  {using {\it Gemini}/T-ReCS} at $18.3\ \micron$ by \citet{churcher11}, with the region interior to $28$~au reported to have little emission. The disk was then detected in scattered light in a re-analysis of the archival 2006 {\it HST}/NICMOS observations by \citet{soummer14}, with the apparent disk extent and orientation consistent with \citet{churcher11}. 

To further characterize the debris disk, we observed the target with {\it HST}/STIS  and {\it Gemini}/GPI. {By carrying out a multi-wavelength study, we aim to understand}: (1) the spatial distribution of the dust; (2) the scattered light color of the dust; (3) the scattering phase functions of the dust for the different spatial components of the system (if available), and how they suit the trends of the current observed debris disks; (4) the dust properties, including size distribution, structure, and compositional information; and (5) whether a universal description of the dust is able to explain the observations across different wavelengths {and observational techniques}.

The structure of this paper is as follows. In Section~\ref{sec-obs}, we describe our HD~191089 observations and the data reduction procedure. In Section~\ref{sec-measure}, we describe {measurables derived from the observations}.  In Section~\ref{sec-diskrt}, we describe our radiative transfer modeling efforts in studying the disk. In Sections \ref{sec-diss} and \ref{sec-conc}, we discuss our findings and provide concluding remarks.

\begin{deluxetable*}{llrrrrrrc}[htb!]
\tablecaption{Observation Log  \label{tab:log}}
\tablehead{
\colhead{Instrument}	& \colhead{Filter}& \colhead{$\lambda_{\rm c}$\tablenotemark{a} } &\colhead{Pixel Scale} & \colhead{IWA\tablenotemark{b} }& \colhead{$T_{\rm exp}$}& \colhead{$N_{\rm frame}$}& \colhead{$\Delta\theta_{\rm PA}$}& \colhead{UT Date}\\
\colhead{}	& \colhead{}& \colhead{($\mu$m)}&\colhead{(mas\,pixel$^{-1}$)} & \colhead{(\arcsec)}& \colhead{(s)}& \colhead{}& \colhead{(\degr)}& \colhead{}}
\startdata
NICMOS		&F110W	&1.12&75.65&0.3	&4607.34	&16	&30.0	& 2006 May 27\\
STIS	&50CCD	&0.58	&50.72&0.3	&2048.00	&64 &109.8	&2014 Jul 19, 2014 Aug 13\\
	&	&&&0.5	&5799.60	&12	&	&\\
GPI			&$H$-Pol	&1.65	&14.166&0.123	&2484.72	&28	&102.2	&2015 Sep 01
\enddata
\tablenotetext{a}{For STIS, $\lambda_{\rm c}$ is the pivot wavelength.}
\tablenotetext{b}{IWA for STIS is the half-width of the wedge-shaped occulter.}
\end{deluxetable*} 

\begin{deluxetable}{lccc}
\tablecaption{Reduction Parameters  \label{tab:reduce}}
\tablehead{
\colhead{}	& \colhead{NICMOS}& \colhead{STIS} & \colhead{GPI}}
\startdata
Classical RDI & N\tablenotemark{a}  & Y\tablenotemark{a} &  N\\
Reference (Classical) &  N & HD~196081 &  N\\
PSF References \# & 85 & 45 (A0.6) & 100 \\
 			&  & 84 (A1.0)  &  \\
KLIP Truncation \# & 19 &  N & 20 \\
NMF Truncation\tablenotemark{b} \# & 10 & 10 (A0.6) & 20 \\
				 &  & 10 (A1.0) &  \\
Polarimetry & N & N & Y \\
\enddata
\tablenotetext{a}{Y: performed; N: not performed or unavailable.}
\tablenotetext{b}{The NMF reductions become stable with more than 10 components.}
\end{deluxetable} 

\section{Observations \& Data Reduction}\label{sec-obs}

\begin{figure*}[hbt!]
\center
\includegraphics[width=\textwidth]{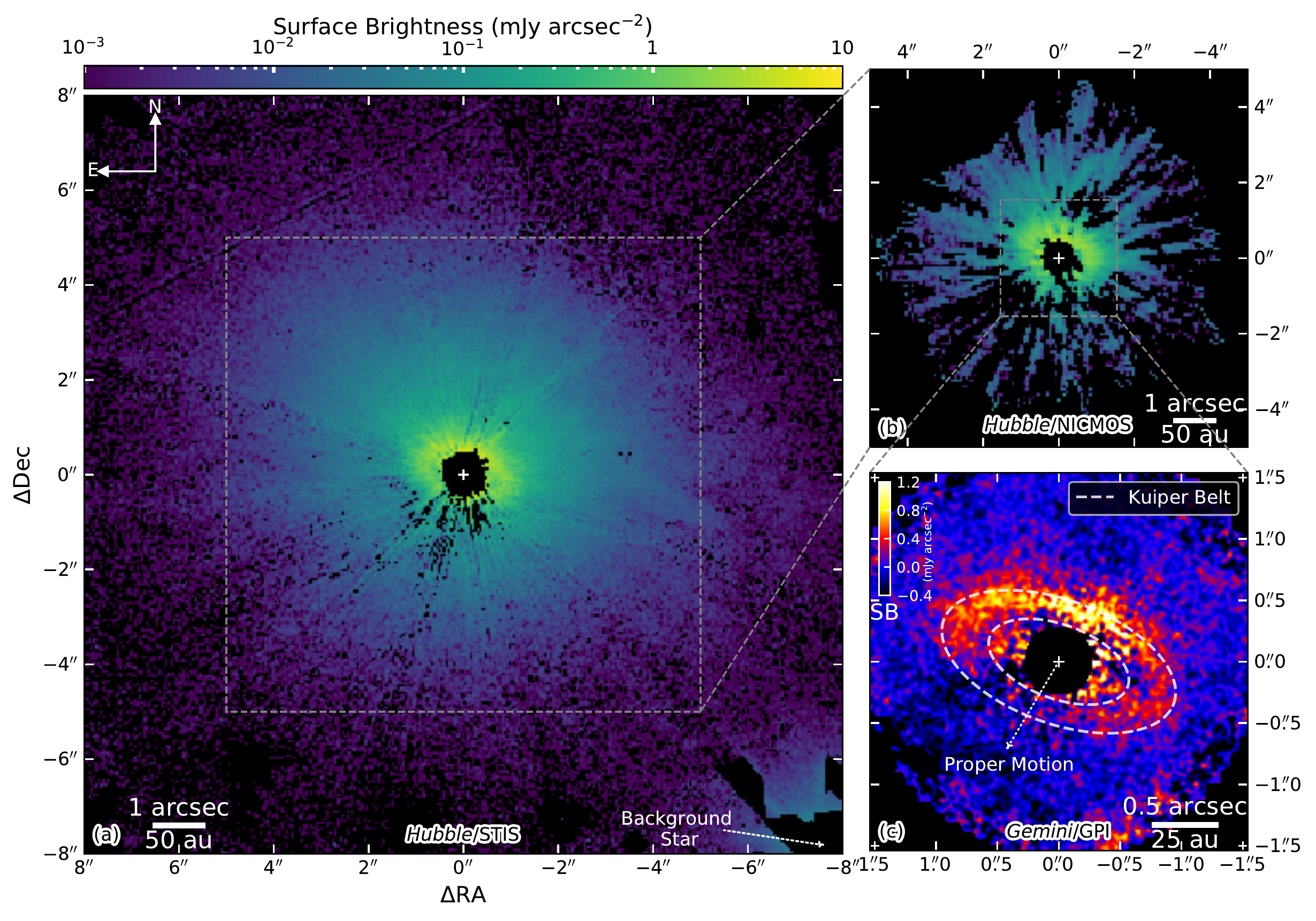}
\caption{2014 {\it HST}/STIS (cRDI, 0.58 $\mu$m), 2006 {\it HST}/NICMOS (NMF, 1.12 $\mu$m), and 2015 {\it Gemini}/GPI (\Qr, 1.65 $\mu$m) images of the HD~191089 debris disk. The outer fan-like structure is unambiguously recovered for the first time with STIS and NICMOS. The GPI \Qr\ image provides the highest spatial resolution. Note: (1) the star positions are marked with white ``+'' signs, and the corresponding signal-to-noise maps are shown in Figure~\ref{fig:SNR}. (2) To mask out regions of significant residual artifacts, a numerical mask with  twice the radius of the GPI mask is used. (3) The units of these images are mJy arcsec$^{-2}$. The STIS and NICMOS data are shown in the same log-scale to better display the halo, while the GPI data are shown in linear-scale to best reveal the ring. (4) For scale, the Solar System Kuiper Belt ($30$--$50$~au, \citealp{stern97}) is illustrated with dashed ellipses on the GPI image, and the proper motion of star HD~191089 is marked with a white, dotted arrow with length corresponding with 10 year motion from {\it Gaia} DR2.}
\label{fig:disk}
\end{figure*}

\subsection{Data Observation and Reduction}\label{sec-obs-data}

\subsubsection{{\it HST}/NICMOS (2006)}\label{sec-hst-nicmos-2006}
HD~191089 was observed using {\it HST}/NICMOS (Proposal ID: 10527, PI: D.~Hines) with the NIC2-CORON aperture and F110W filter ($\lambda_{\text{c}}=1.12\ \micron$, inner working angle [IWA]: $0\farcs3$, pixel scale: $75.65$~mas~pixel$^{-1}$, \citealp{nicmosbook09}) on 2006 May 27, with two telescope orientations each observing the target for $2303.67$ s. These {\it HST}/NICMOS data, totaling 16 frames, have previously been presented in \citet{soummer14}. We perform an updated reduction including application of two different point spread function (PSF) subtraction algorithms to remove the star light and speckle noise and {reveal} the debris disk around HD~191089.

We obtain the scattered light image of the system using the Multi Reference Differential Imaging (MRDI) technique. {Specifically, we retrieve 849 F110W exposures} of 70 diskless reference stars in the ALICE archive of the NICMOS observations (PI: R.~Soummer, \citealp{choquet14, hagan18})\footnote{\url{https://archive.stsci.edu/prepds/alice/}}. For each target exposure, we  {first} decompose its 10\% closest ALICE images in correlation (i.e., $L^2$-norm sense {, and a total of 85 images}) using both the Karhunen-Lo\`eve Image Projection (KLIP, \citealp{soummer12}) and the Non-negative Matrix Factorization (NMF\footnote{ {Python-based {\tt nmf\_imaging} package: \citet{nmfimaging}.}}, \citealp{ren18}) data reduction methods, then model the target with these components. The disk then resides in the residual image when the empirical PSF model is subtracted from the target exposure. The final NICMOS disk image is then the element-wise mean of the derotated individual residual exposures\footnote{The ${\sim}0.9\%$ $x$- and $y$-direction scale difference in \citet{schneider03} is ignored.}.

To calibrate the NICMOS disk image and obtain a measurement of its surface brightness, we multiply the reduced data by the calibrated F110W {\tt PHOTFNU} parameter\footnote{\url{http://www.stsci.edu/hst/nicmos/performance/photometry/postncs_keywords.html}} $F_\nu=1.21\times 10^{-6}$ Jy\,s\,count$^{-1}$, then divide the data by the NICMOS pixel area on-sky to obtain the surface brightness data in units of Jy\,arcsec$^{-2}$.

\subsubsection{{\it HST}/STIS (2014)}
HD~191089 was observed using {\it HST}/STIS (Program ID: 13381, PI: M.~Perrin) using the 50CORON aperture ($\lambda_{\text{c}}=0.58\ \micron$, pixel scale: $50.72$~mas~pixel$^{-1}$, \citealp{stisihb18}) on 2014 July 19 and 2014 August 13 (2 visits each), totaling 76 frames. Each visit was performed with a different telescope orientation, with position angles of $-84\fdg23, -61\fdg23, -2\fdg23$, and $25\fdg59$ for the $y$-axis in the images (N to E). These telescope orientations are selected to obtain 360\degr\ azimuthal coverage of the disk down to the occulting mask (similar to \citealp{schneider14}). In each visit, we first obtained 16 short $32$~s exposures on the WEDGEA0.6 position to probe the inner region down to a half-width of $0\farcs3$. Then we obtained 3 longer 483.3~s exposures on the WEDGEA1.0 position (half-width: $0\farcs5$) to deeply probe the exterior region of the disk.

To perform PSF subtraction, we also observed the reference star HD~196081 (selected for color, brightness, and on-sky proximity matches to HD~191089\footnote{\url{http://www.stsci.edu/hst/phase2-public/13381.pro}}) using the same aperture positions, with one visit interleaved between the two science (HD~191089) visits at each epoch. With the numerical mask created by \citet{debes17} to mask out the STIS occulters, we perform multiple PSF subtractions using different approaches: We first apply the classical Reference Differential Imaging (cRDI) technique to subtract the star light in each exposure by minimizing the residual variation in the region of the coronagraphically unapodized diffraction spikes (excluding where the disk resides). We also obtain the STIS reference exposures from the STIS PSF archive created in \citet{ren17} for MRDI reduction: for each target exposure, we perform NMF reduction using the 10\% most correlated references in the STIS archive (i.e., 85 images for WEDGEA1.0, and 45 images for WEDGEA0.6). The final STIS disk image is then the element-wise mean of the individual derotated PSF-subtracted exposures.

To calibrate the STIS image in physical surface brightness units, we convert the {\tt PHOTFLAM} photometric parameter in the raw FITS file for the HD~191089 observations ($F_{\lambda}=4.15\times 10^{-19}$~erg\,cm$^{-2}$\,\AA$^{-1}$\,count$^{-1}$) to $F_\nu=4.56\times 10^{-7}$~Jy\,s\,count$^{-1}$ using the conversion equation in Appendix B.2.1 of \citet{nicmosbook09}: $$F_\nu = \frac{\lambda_{\rm c}^2F_\lambda}{3\times10^{-13}},$$ where $\lambda_{\rm c}=0.58\ \mu$m is the pivot wavelength of STIS. We then multiply our combined STIS image in count~s$^{-1}$ pixel$^{-1}$ by $F_\nu$, and divide it by the STIS pixel area on-sky to obtain the surface brightness data in units of Jy\,arcsec$^{-2}$.

\subsubsection{{\it Gemini}/GPI (2015)}
HD~191089 was observed using {\it Gemini}/GPI in $H$-band ($\lambda_{\text{c}}=1.65\ \micron$, pixel scale: $14.166\pm0.007$ mas~pixel$^{-1}$, \citealp{derosa15}) polarimetric mode (``$H$-Pol'', \citealp{perrin15})  on  2015 September 01  during the Gemini Planet Imager Exoplanet Survey (GPIES; PI: B.~Macintosh, \citealp{macintosh14}). We took 28 exposures, each with $88.74$ s integration time with a total field rotation of $102\fdg2$. The airmass ranged from $1.002$ to $1.01$, the differential motion image monitoring (DIMM) seeing measurement was $1\farcs16\pm0\farcs17$, and the multi-aperture scintillation sensor (MASS) seeing was $1\farcs1\pm0\farcs3$. 

To obtain the Stokes cube (\{$I, Q, U, V$\}) for the HD~191089 debris disk, we follow the recipes described in \citet{perrin14} and \citet{millarblanchaer15}, and reduced the raw exposures using the GPI Data Reduction Pipeline \citep[DRP,][]{perrin14, perrin16} and the~automated data processing architecture \citep[Data Cruncher:][]{wang18}. The $Q$ and $U$ components in the  {traditional} Stokes cube were then transformed to the local Stokes cube (\{\Qr, \Ur\}), with \Qr\ representing the polarized light perpendicular or parallel to the radial direction, and \Ur\ at ${\pm}45^\circ$ from it \citep{monnier19}.  On the local Stokes maps of HD~101089, we notice two similar low spatial frequency octopole structures with a rotation of ${\sim}45^\circ$. However, for a \Ur\ map, we do not expect any signal from an optically thin disk with single scattering events on the dust (see \citealp{canovas15} for a discussion of the validity and exceptions). Given that we observe similar structures in the other GPI polarimetry observations, we expect such a structure is to be one of the instrumental artifacts \citep{esposito19prep}. To reduce the systematic errors induced by it, we fit an octopole model using the \Ur\ map and remove it from the \Ur\ map, then rotate the model $45^\circ$ and remove it from the \Qr\ map.

We flux calibrate these data
following the procedure described in \citet{hung15, hung16}: we first
adopt for HD~191089 an $H$-band flux of $F_\star=3.749\pm0.119\,\rm{Jy}$ \citep{2mass06}, then adopt the satellite-to-star ratio $R=2.035\times10^{-4}$ from GPI DRP \citep{wang14, perrin14} and average satellite spot  {total} flux $S=(1.12\pm0.17)\times10^3\, \rm{count\,  s^{-1}}$ in the FITS file header. Combining these, we obtain a conversion factor of
\begin{equation*}
F_\nu=\frac{RF_\star}{S}=(6.87\pm1.06)\times10^{-7}\, \rm{Jy\, s\, count^{-1}}.
\end{equation*}

We apply that conversion factor and normalize  {the local Stokes maps} by the exposure time and pixel area to obtain the disk surface brightness data in units of Jy\,arcsec$^{-2}$. The flux-calibrated data are then geometrically corrected and smoothed by convolving with a Gaussian kernel ($\sigma=14.166$~mas, i.e., the scale of one GPI pixel: \citealp{millarblanchaer16}) to remove the high spatial frequency noise that impacts regions smaller than the Nyquist-sampled point-source PSF of GPI.
\subsection{Noise Estimation}\label{sec-obs-noise}

\begin{figure*}[htb!]
\center
\includegraphics[width=\textwidth]{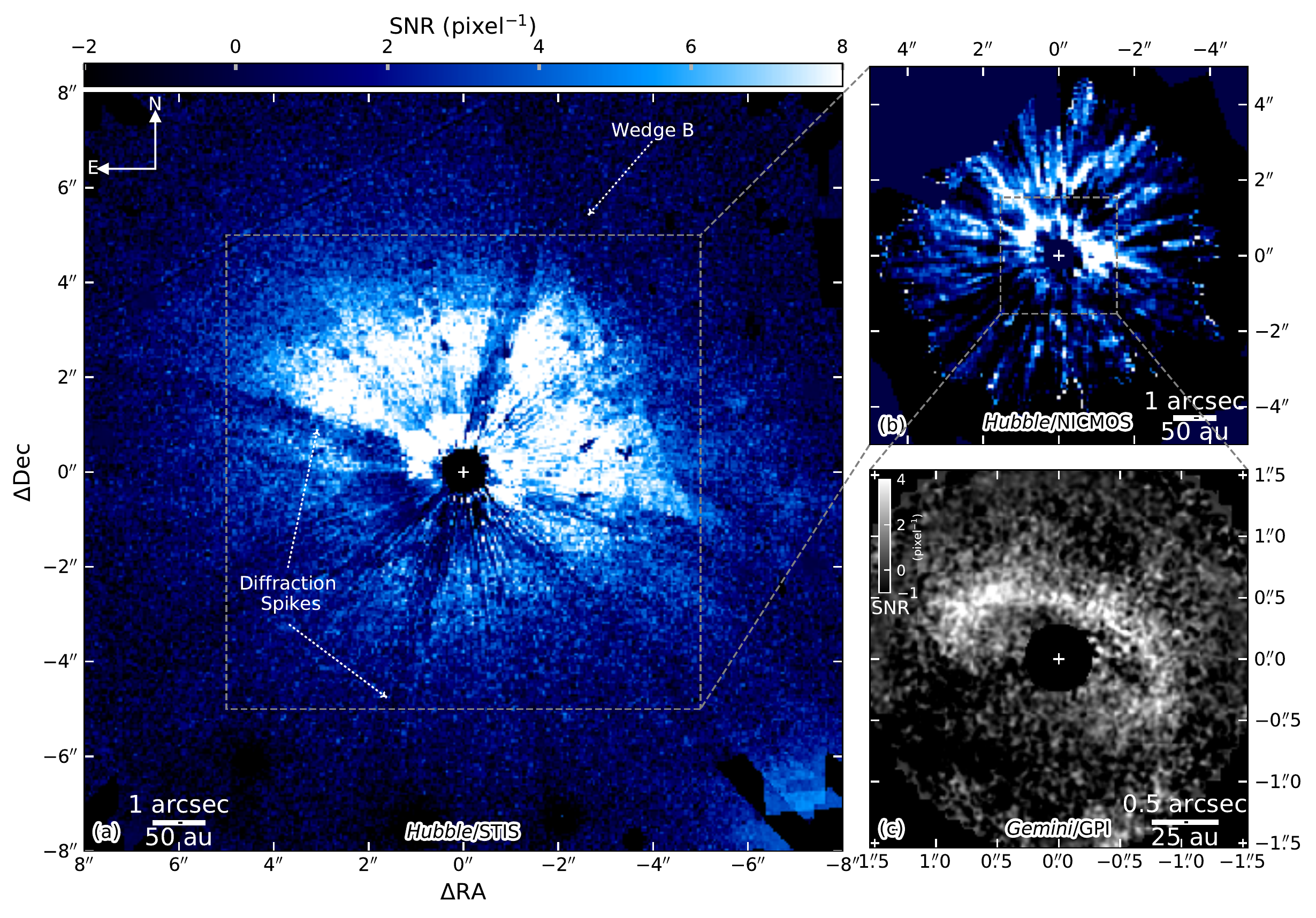}
\caption{SNR maps of the reduction results in Figure~\ref{fig:disk}. In the STIS SNR map, the presence of Wedge B truncates the halo on the north-west side, and the coronagraphically unapodized diffraction spikes reduce the SNR. Note: (1) the STIS and NICMOS data are shown in the same scale. (2) Unless otherwise specified, the correlated noises are not estimated in this paper.  {(3) The instrument pixel scales are illustrated by black dashes at the center of the scale bars.}}
\label{fig:SNR}
\end{figure*}

Based on different PSF subtraction methods for each dataset, we estimate the uncertainties as follows.

{\it NMF \& cRDI} (STIS, NICMOS): We estimate the noise in these reductions from the ensemble of science frames to probe the temporal variations from frame to frame (16 NICMOS frames, 64 and 12 STIS frames at the two different positions on the mask, see Table~\ref{tab:log}). We proceed as follows: after subtraction of the PSF, we compute the  {pixel-wise} standard deviation across the science frames to obtain the typical noise map per frame. which is used to account for the noise added by PSF subtraction. We then replicate this noise map for $N_{\rm frame}$ times, and derotate each with the same angle as each science frame. We obtain the final noise map by computing the square root of the quadratic sum of these derotated noise maps.

{\it KLIP} (NICMOS): For the NICMOS-KLIP reduction, we use the ALICE library of reference stars that are processed identically to HD~191089 to estimate the residual speckle noise. For each HD~191089 image (16 total), we first select $25\%$ of the most correlated images in the reference library (i.e., $212$ reference images). In this way, we obtain 413 reference images that are correlated with at least one HD~191089 image (i.e., ${\sim}50\%$ of the entire library). We split these $413$ images into 25 groups each containing 16 images (with the left-overs randomly discarded). In each group, the $16$ images are treated as the mock target images to simulate 16 HD~191089 non-detection images. For one mock target image, we first identify the real target that was observed, then remove the images that are taken on the same real target from the reference library to avoid self-subtraction. We then use the updated reference library to perform KLIP subtraction of the mock image following the identical procedure as for HD~191089 (i.e., 19 eigenmodes from 84 most correlated references). The 16 reduced mock images are then derotated using the same orientation angles as the ones in the HD~191089 observations. For each group, we take the element-wise mean of the 16 rotated reduced mock target images as the mock result for one realization of HD~191089 non-detection. We then obtain a total of $25$ realizations of non-detections using the $25$ full groups from the $413$ mock images. We take the element-wise standard deviation from the $25$ mock non-detections as the noise map for our NICMOS-KLIP reduction.

\begin{figure}[htb!]
\center
\includegraphics[width=0.5\textwidth]{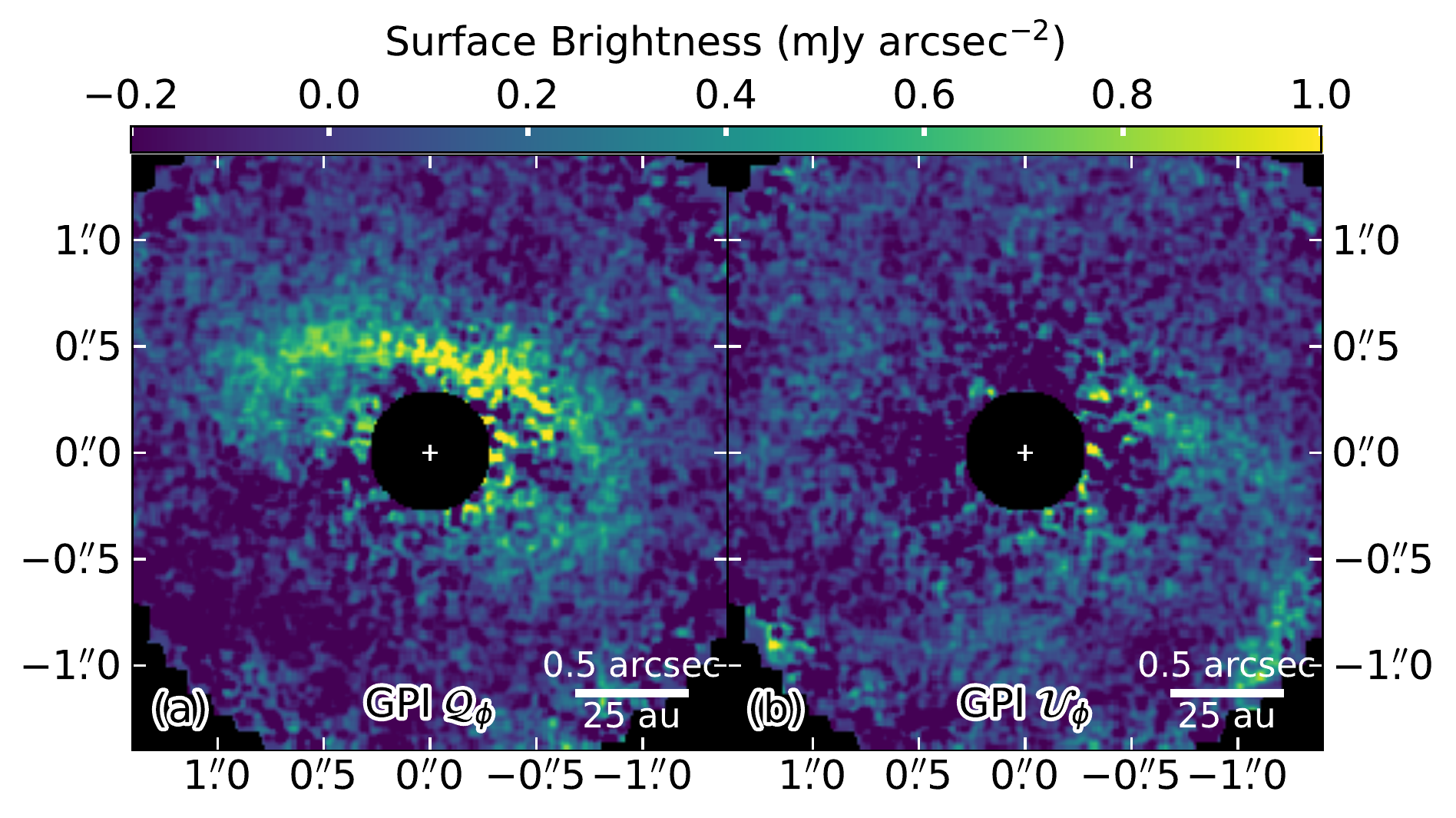}
\caption{Smoothed and octopole-removed GPI $H$-band (a) \Qr\ and (b) \Ur\ maps.}\label{fig:image-gpi}
\end{figure}

{\it Polarimetry} (GPI): For the GPI \Qr\ map, we used the  {convolved} \Ur\ image as a proxy for the uncertainty \citep{millarblanchaer15}.  {To obtain the noise map, for each angular separation to the star, we calculate the standard deviation of an annulus with 3 pixel width in the convolved \Ur\ image as the noise.} This noise map provides a reasonable estimation of PSF subtraction residuals, photon and detector noise, and residual instrumental polarization \citep{millarblanchaer15}. See Figure~\ref{fig:image-gpi} for the convolved \Qr\ and \Ur\ images used for SNR calculation.

\subsection{Comparison of the Reduction Methods}\label{sec-obs-compare}
\begin{figure*}[htb!]
\center
\includegraphics[width=0.98\textwidth]{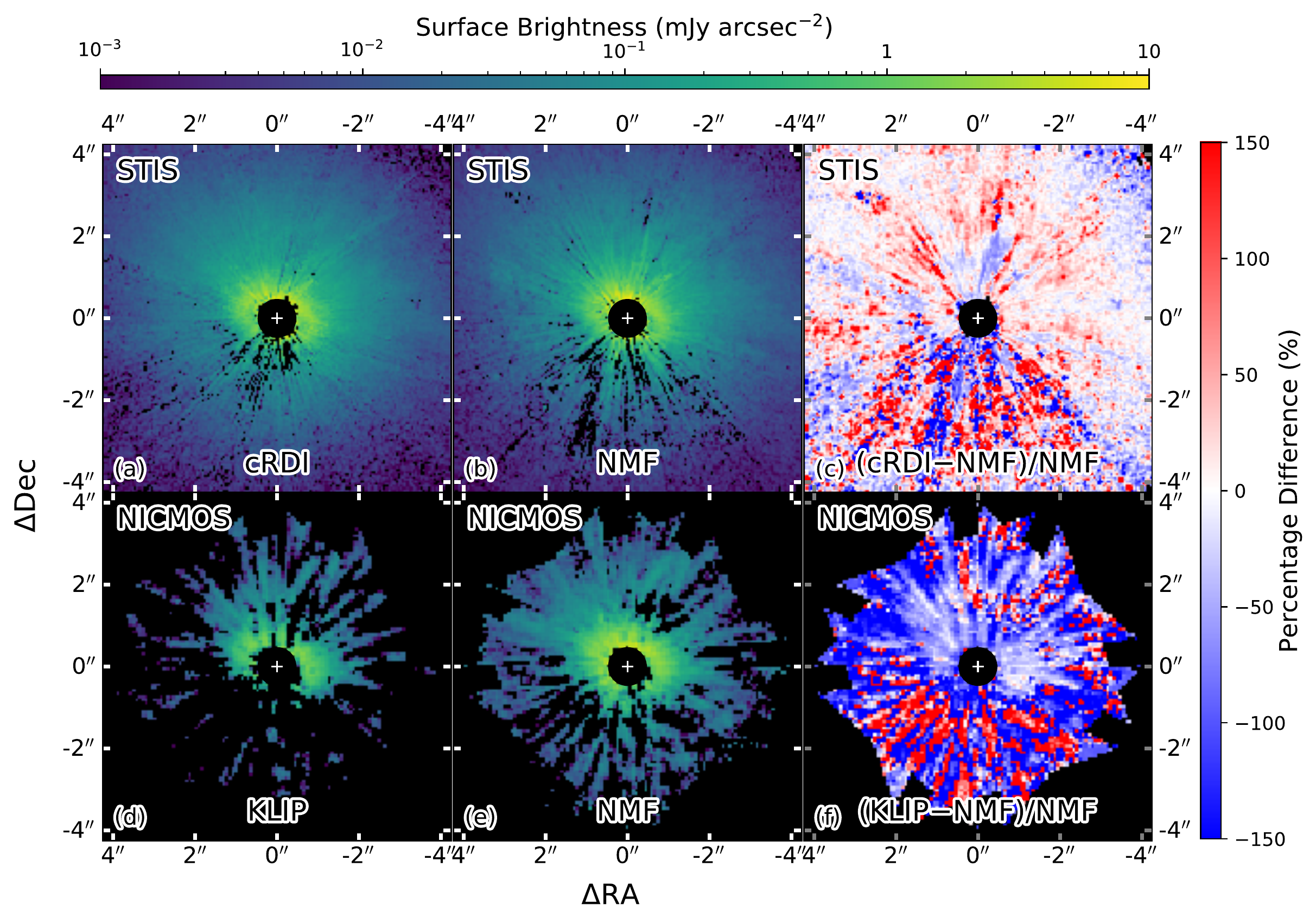}
\caption{Comparison of the HD~191089 ring between the NMF reduction (middle) and the other reduction results (left). The right column shows the percentage difference from the NMF image when it is subtracted from the image on the left column. For the STIS data (top), the classical RDI and NMF image photometry agree to within ${\sim}10\%$ per pixel for the ring; for the NICMOS data (bottom), NMF recovers nearly twice the disk flux recovered by KLIP. See Figure~\ref{fig:SNR-methods} for the SNR maps.}\label{fig:image-methods}
\end{figure*}

\begin{figure}[htb!]
\center
\includegraphics[width=.52\textwidth]{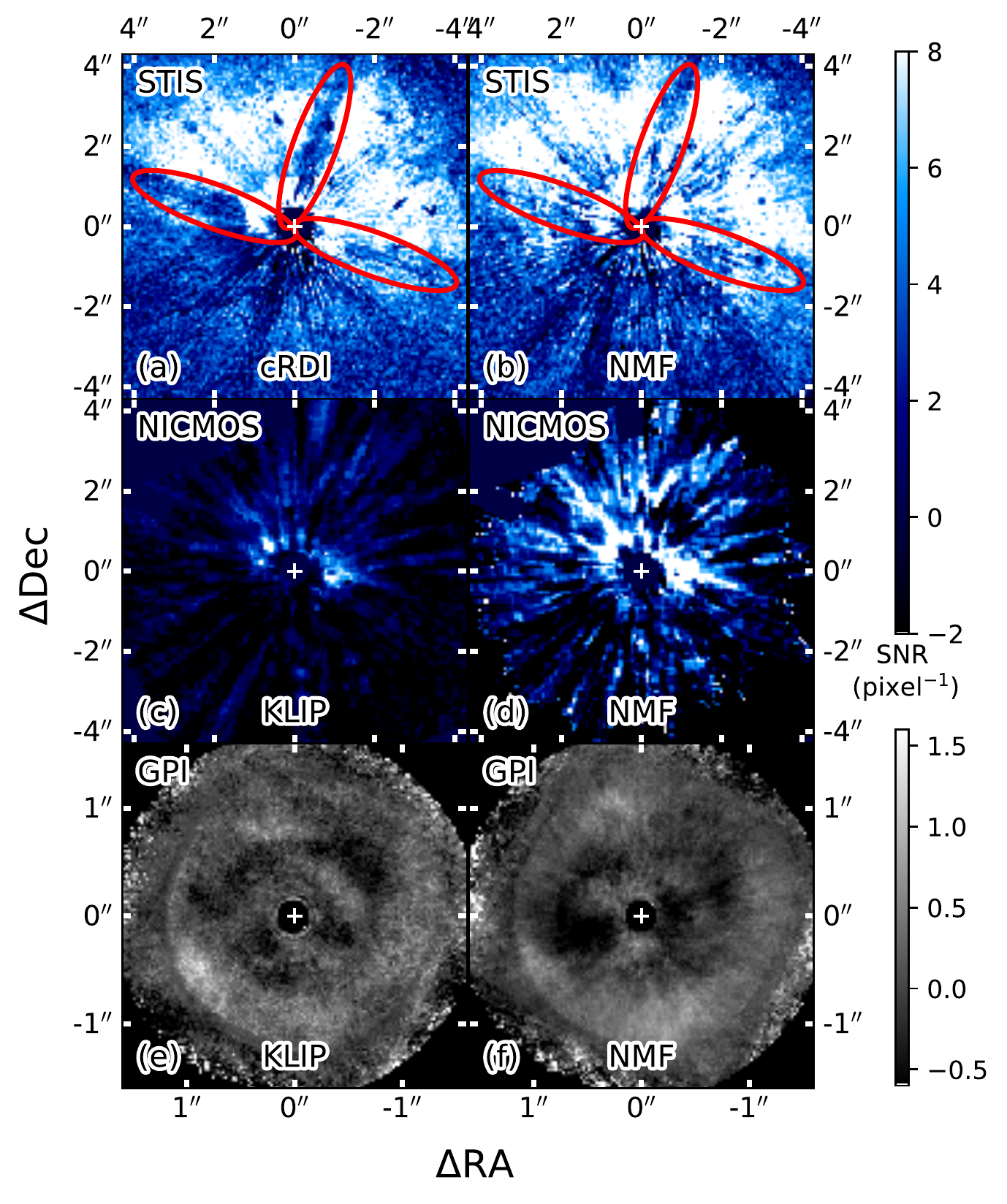}
\caption{SNR maps of the reduction results with different methods. For the STIS image, both cRDI and NMF reach similar SNR levels {, and the regions marked by red ellipses in NMF have  ${\sim}3$ times the SNR in the cRDI result}. For the NICMOS image, NMF is able to recover higher SNR than KLIP. For the GPI image, neither of the methods are able to extract the disk structure.}\label{fig:SNR-methods}
\end{figure}
The HD~191089 debris disk is detected in all of our {\it HST} and GPI observations. We here compare the quality of the different reductions and discuss their relative merits for measuring different quantities of interest.\\

{\it STIS}: The disk is detected in the STIS data with  high morphological and photometric fidelity in both the cRDI and the NMF reductions (Figure~\ref{fig:image-methods}). The system shows a bright parent belt, surrounded by a faint and diffuse halo detected up to ${\sim}6\arcsec$ from the star. These features are detected with a dynamic range of 3 orders of magnitude, with the {peak ring surface brightness of} ${\sim}1$ mJy\,arcsec$^{-2}$ and the diffuse halo  {detected} down to a few $\mu$Jy\,arcsec$^{-2}$. Theoretically, unlike KLIP reduction, both cRDI and NMF reduction methods are free from over-subtraction caused by over-fitting disk features using the references. This was discussed in \citet{ren18} and is confirmed in \citet{ren18b} where the NMF method is able to successfully retrieve the spiral arms for the MWC~758 system. 

For HD~191089, the bright disk and excellent cRDI quality enables a quantitative comparison between cRDI and NMF. In Figure~\ref{fig:image-methods}, we present the reduction results from different methods for both the STIS and the NICMOS data, for a comparison between the methods. For the STIS data, the ring surface brightness is consistent within $\pm10\%$ between cRDI and NMF. NMF achieves higher SNRs along the major and minor axes of the disk  {by a factor of ${\sim}3$}, i.e., the regions containing coronagraphically unapodized diffraction spikes {, which are marked by red ellipses} in Figure~\ref{fig:SNR-methods}.\\

{\it NICMOS}: The halo component discovered in STIS is confirmed in the NICMOS-NMF image. The NICMOS-KLIP image in Figure~\ref{fig:image-methods} is able to recover the halo that was not observed in the original discovery image \citep{soummer14}. The new reduction is obtained with a larger field of view and fewer KLIP components.  In retrospect, it is not surprising that the original reduction by \citet{soummer14} did not detect the halo: the KLIP method is based on principal component analysis, which requires mean-subtraction for each individual image. When modeling the target  {with KLIP}, the reduced image also has zero mean, which offsets the faint halo as negative background, thus the halo cannot be recovered with positive signals.  {In addition,} to maximize the removal of the star light, a large number of KLIP components was used in \citet{soummer14}. In this case, KLIP also removed the halo because its extended diffuse structure resembles PSF wing.
 
NMF  {independently} confirms the existence of the  halo in the NICMOS data while recovering the ring regardless of increasing component number. The NMF component basis is non-negative --- thus, not orthogonal --- and it does not perform direct projection of vectors as KLIP which falls into the overfitting regime, but instead searches for the best non-negative combination of non-negative NMF components. The halo does not resemble the NMF components that are used to model the PSF wings, and thus it remains in the residual after PSF subtraction.

In the NICMOS results in Figure~\ref{fig:image-methods}, the NIMCOS-NMF image has ${\sim}2$ times the ring surface brightness of the NICMOS-KLIP image, supporting the expected behavior of the two methods: KLIP's over-fitting because of direct vector projection cannot be avoided even with a small number of eigen-modes, and the mean-subtraction offset reduces the overall flux of the system. In contrast, NMF is expected to preserve the surface brightness of the NICMOS disk as for the STIS data. 
\\

{\it GPI in Total Intensity}: We attempted to detect the disk in total intensity from this same dataset with MRDI: for each polarization-direction pair of $H$-Pol exposures, we derive a single total intensity image. For starlight subtraction, we select the  $100$ most correlated GPI $H$-band PSFs from the library of $15,847$ exposures, then perform KLIP and NMF reductions. We present the SNR maps of the null-detections in Figure~\ref{fig:SNR-methods}: {we do not detect the disk or any point source using either KLIP or NMF.} {We hypothesize that} a larger or better reference library is needed to obtain the best match for HD~191089. 

 {To estimate an upperlimit on the surface brightness for the disk, we calculate} the noise map for the GPI-KLIP image using the standard deviation across the individual reduced images. {We estimate a} $1\sigma$ uncertainty of ${\sim}3$ mJy arcsec$^{-2}$ at the minor axis, and ${\sim}1$ mJy arcsec$^{-2}$ along the major axis. Assuming gray scattering for the dust in the ring seen with the NICMOS F110W filter and GPI $H$-band, the NICMOS-NMF result will produce surface brightness of ${\sim}2.2$ mJy arcsec$^{-2}$ and ${\sim}1.0$ mJy arcsec$^{-2}$, respectively. Therefore, even the disk is not removed by the reduction methods, it is below a detection threshold of $1\sigma$ (which corresponds to a total SNR of 5 for extended structure spanning ${\sim}25$ pixels: \citealp{debes19}). In this way, its detectability is beyond the limit of the current methods with the current GPI $H$-band PSF library.

\section{Disk Morphology \& Measurement}\label{sec-measure}

\subsection{Strategy to Measure Disk and Dust Properties}\label{sec-instrument-compare}
For the ring, the GPI \Qr\ map {provides} the highest spatial sampling, with STIS and NICMOS offering total intensity observations at different wavelengths. Using the GPI \Qr\ map, we obtain the geometric structure for the ring in Section~\ref{sec:ring-analyzer}, including inclination, semimajor axis, position angle, {ring center position, as well as the eccentricity of the deprojected ring}. With the geometric information, and the parallactic distance to HD~191089, we are able to calculate the average intensity of the disk as a function of physical distance to the host star (i.e., the radial profile).

For the halo, the STIS total intensity image is able to probe the largest spatial extent  {with larger field of view and high sensitivity}. Assuming that the halo is coplanar with the ring, we can measure the radial profile for the halo, which will help to identify {whether the halo is a geometric extension of the ring by comparing their outer power law indices}. We use the cRDI reduction for this purpose, since it covers a larger field of view than the NMF reduction obtained from the fixed-width STIS archive in \citet{ren17}.

For both the ring and the halo, if we assume they are coplanar, then we can measure the disk surface brightness as a function of scattering angle---the scattering phase function (SPF). The SPF is related to properties of the dust and therefore provides insights into the composition, size distribution, and minimum dust size for the system even though this information is degenerate. To extract the information for the dust, we {adopt the spatial distribution information from GPI \Qr\ measurements}, then use radiative transfer modeling tools to model the ring. Given that the ring is well resolved with all three instruments, the SPFs {at different wavelengths are expected to} help constrain the dust size and composition.

\subsection{Mathematical Description}\label{sec-describe}
 {The spatial distribution of the dust in debris disk can be parameterized using} a 3-dimensional function in cylindrical coordinates: the radial distribution in the mid-plane, and the vertical distribution along the axis perpendicular to the mid-plane. Along the radial direction the dust in the system follow a combination of two power laws
\begin{equation}\label{eq-powerlaw}
\rho(r)\propto \left\{\left(\frac{r}{r_{\rm c}}\right)^{-2\alpha_{\rm in}}+\left(\frac{r}{r_{\rm c}}\right)^{-2\alpha_{\rm out}} \right\}^{-\frac{1}{2}},
\end{equation}
where $\alpha_{\rm in}>0$ and $\alpha_{\rm out}<0$ approximate the {\it mid-plane} dust density power law indices interior and exterior to $r=r_{\rm c}$ \citep{augereau99}. Along the normal direction of the disk mid-plane, the dust follows a Gaussian dispersion form, i.e., 
\begin{equation}\label{eq-z}
Z(r, z) \propto \exp\left[-\left(\frac{z}{\zeta(r)}\right)^2 \right],
\end{equation}
with \begin{equation}\label{eq-flare}
\zeta(r) = h r^\beta,
\end{equation}
where \begin{equation}\beta=1\end{equation} for a non-flared debris disk. {For} the HD~191089 system, $h=0.04$ is adopted from the vertical structure study {\citep{thebault09}}. 

The the power law index for the surface density radial profile is then \citep{augereau99},\begin{equation}
\Gamma = \alpha + \beta,\label{eq-gamma}
\end{equation}
where $\alpha$ is $\alpha_{\rm in}$ or $\alpha_{\rm out}$ in Equation~\eqref{eq-powerlaw}  {, with $\Gamma$ being $\Gamma_{\rm in}$ or $\Gamma_{\rm out}$ as the corresponding indices}.

{Since illumination decreases as a function of distance from the star, our images must be corrected for illumination effects before the surface density can be measured. The relationship between the surface brightness power law index, $\gamma$, and the surface density power law index, $\Gamma$, is
\begin{equation}\label{eq-sb}
\gamma = \Gamma - 2,
\end{equation}
where $\gamma$ is $\gamma_{\rm in}$ or $\gamma_{\rm out}$, corresponding with $\Gamma$ being $\Gamma_{\rm in}$ or $\Gamma_{\rm out}$.}

The radial distribution of the system in Equation \eqref{eq-powerlaw} is supplemented with two extra parameters $r_{\rm in}$ and $r_{\rm out}$, which are the cutoff radii. They are introduced to describe the clearing of materials interior and exterior to the disk, i.e., when $r < r_{\rm in}$ or $r > r_{\rm out}$, $\rho(r) = 0$.

\subsection{Disk Morphology}\label{sec:disk-morph}
\subsubsection{Ellipse Parameters}\label{sec:ring-analyzer}
We measure the geometric parameters for the disk from the GPI \Qr\ observations, because it has the smallest pixels, and is less biased by the reduction methods. 

We assume that the peak radial polarized surface density matches the peak radial particle density, and use the {\tt Debris Ring Analyzer} package from \citet{stark14} to fit the peak intensity of the ring in $10^\circ$ azimuthal wedges {by minimizing the $\chi^2$ value between the observation and an ellipse model, and quantify the uncertainties by assessing the change of the $\chi^2$ values for the corresponding degrees of freedom on a grid of the explored parameters \citep{choquet18}}. 

We measure an inclination\footnote{The uncertainties calculated in this paper are $1\sigma$ unless otherwise specified.} of $\theta_{\rm inc}={59^\circ}_{-2^\circ}^{+4^\circ}$ from face-on, {and} a semimajor axis of $45_{-1}^{+2}$~au {, and} a position angle of $\theta_{\rm PA}={70^\circ}_{-3^\circ}^{+4^\circ}$ from North to East  {for the major axis}. There is no significant offset between the location of the star and the center of the ring ($3\sigma$ upper limit: 8~au){. The} $3\sigma$ upper limit for the eccentricity of the deprojected ellipse is $0.3$.

\begin{deluxetable}{cccp{2.7cm}}
\tablecaption{Disk Morphology Parameters  \label{tab:disk-measure}}
\tablehead{
\colhead{Parameter}	& \colhead{Ring} & \colhead{Halo} & \colhead{Meaning} \\ \hline
\colhead{Instrument}	& \colhead{GPI} & \colhead{STIS} & \colhead{}}
\startdata
$\theta_{\rm inc}$\tablenotemark{a}	&	${59^\circ}_{-2^\circ}^{+4^\circ}$		&	$59^\circ$ \tablenotemark{c} 		& Disk mid-plane inclination from face-on.		\\
$\theta_{\rm PA}$\tablenotemark{a}	&	${70^\circ}_{-3^\circ}^{+4^\circ}$		&	$70^\circ$ \tablenotemark{c} 		& Position angle for major axis (N to E).		\\ \hline
$\gamma_{\rm in}$\tablenotemark{b}	&	$4.9{\pm}0.2$		&	$\cdots$\tablenotemark{d} 	& \multirow{2}{2.45cm}{Surface brightness power law indices.}			\\
$\gamma_{\rm out}$\tablenotemark{b}	&	$-6.1{\pm}0.2$		&	$-2.68{\pm}0.04$		&  \\ 
$r_{\rm c}$\tablenotemark{b}		&	$43.6{\pm}0.3$~au	&	$\cdots$\tablenotemark{d} 	& $r_{\rm c}$ in Equation~\eqref{eq-powerlaw}.		\\
$r_{\rm in}$\tablenotemark{b}		&	$26{\pm}4$~au		&	$\cdots$\tablenotemark{d} 	& \multirow{2}{2.45cm}{Inner and outer clearing radii.}			\\
$r_{\rm out}$\tablenotemark{b}		&	$78{\pm}14$~au		&	$640{\pm}130$~au	&	\\ \hline
$r_{\rm center}$& 	$45.6{\pm}0.2$~au 	& 	$\cdots$\tablenotemark{d}  & \multirow{2}{2.45cm}{Peak position and FWHM of a Gaussian ring.}\\
FWHM& 	$24.9{\pm}0.4$~au 	& 	$\cdots$\tablenotemark{d} & 
\enddata
\tablenotetext{a}{Fitted with {\tt Debris Ring Analyzer} \citep{stark14}.}
\tablenotetext{b}{Fitted for Equations~\eqref{eq-powerlaw} \& \eqref{eq-sb}.}
\tablenotetext{c}{Values adopted from the GPI results.}
\tablenotetext{d}{Unconstrained from the STIS image.}
\end{deluxetable}

\begin{figure*}[htb!]
\center
\includegraphics[width=.95\textwidth]{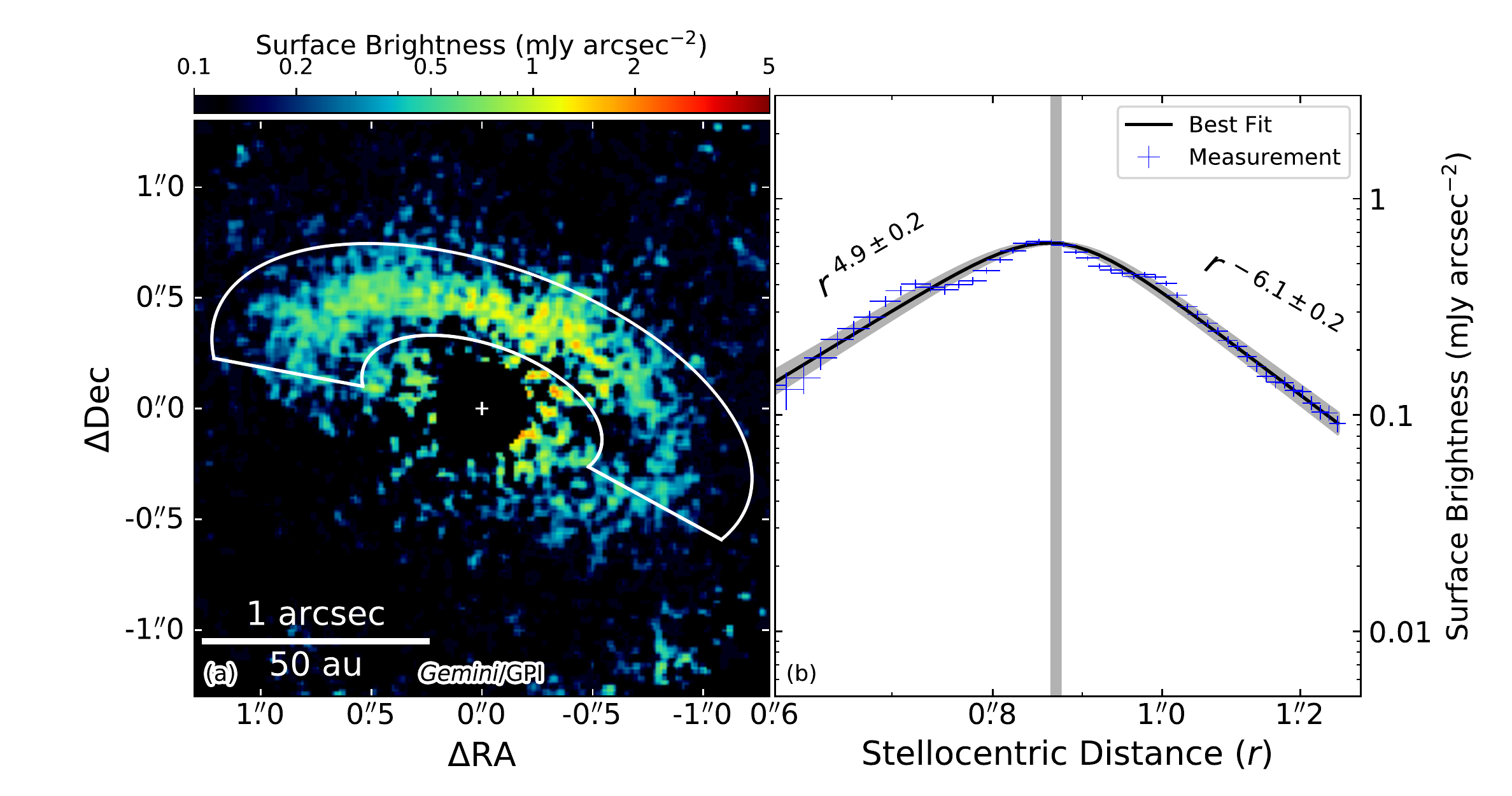}
\includegraphics[width=.95\textwidth]{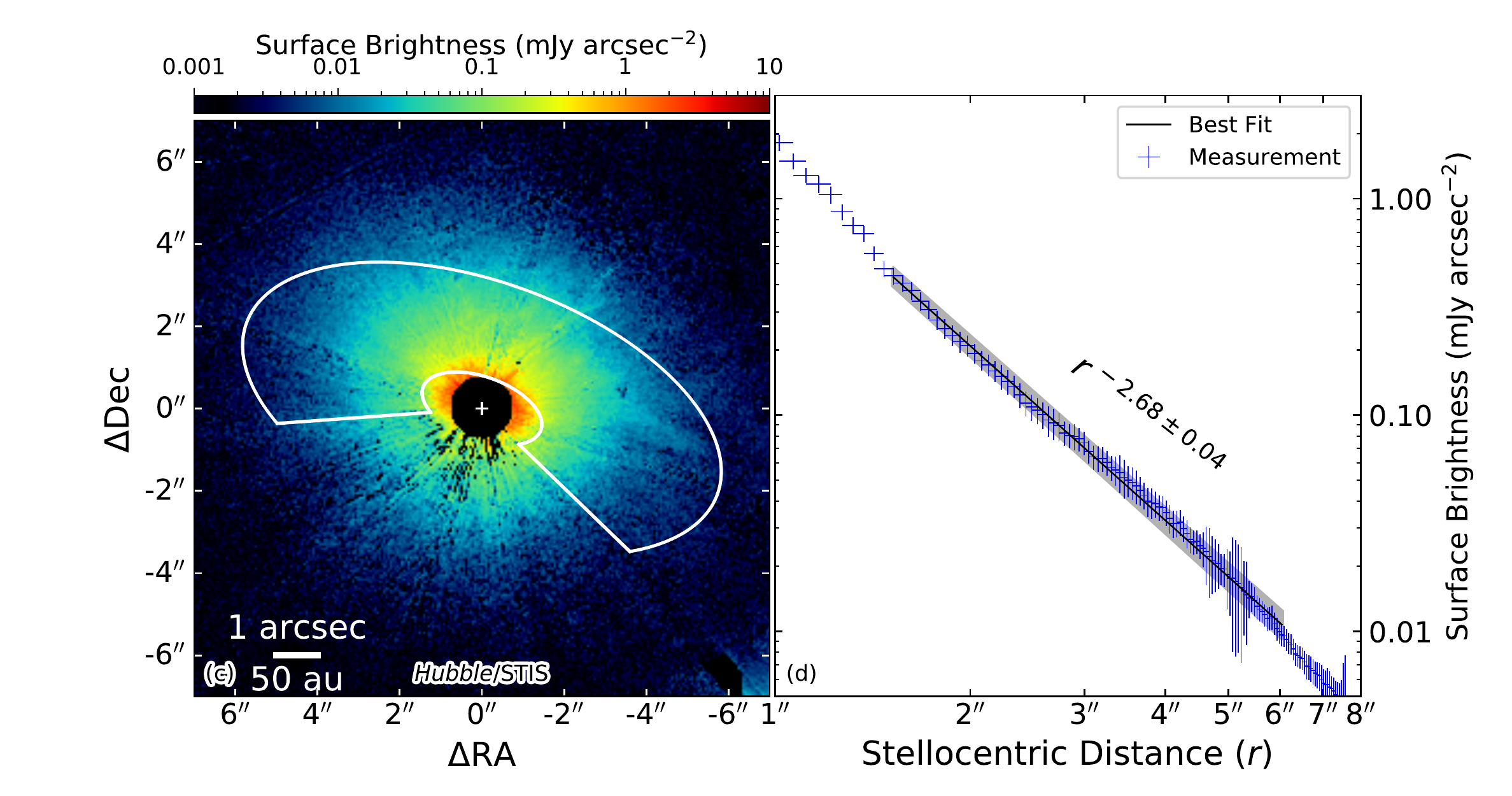}
\caption{{\it Left}: Disk images. {\it Right}: Average flux density radial profiles, and power-law fits for the regions enclosed by white lines  {(uniformly sampled in stellocentric distances)}. For both panel ({\bf b}) and ({\bf d}), the shaded areas are the 1$\sigma$ intervals corresponding with the fitted parameters ($\gamma_{\rm in}$, $\gamma_{\rm out}$, and $r_{\rm c}$). The $\gamma_{\rm out}$ parameter for the ring and the halo differs by more than $10\sigma$, strongly indicating the different spatial distributions of the two components.}
\label{fig:radialprofile}
\end{figure*}
\subsubsection{Radial Distribution}\label{sec-disk-radial-profile}

We measure the surface brightness power law indices by averaging the flux density at the same stellocentric radius in the highest SNR regions.  Although this approach implicitly assumes isotropic scattering, which is a likely incorrect assumption, we tested it by performing the fit in narrow wedges (e.g., along the minor or major axis) and obtained consistent results, indicating that the scattering anisotropy does not change our results significantly. We plot the surface brightness radial profiles for the ring and the halo in Figure~\ref{fig:radialprofile}, and summarize the results in Table~\ref{tab:disk-measure}. For both the ring and the halo, {we define the inner and outer} cutoff radii, $r_{\rm in}$ and $r_{\rm out}$, {as the radii for which the corresponding average surface brightness is} consistent with zero at the $1\sigma$ level. 

We measure the following parameters for the ring from the GPI \Qr\ observation: $\gamma_{\rm in} = 4.9\pm0.2$, $r_{\rm c} = 43.6\pm0.3$~au, and $\gamma_{\rm out}=-6.1\pm0.2$.\footnote{For the input data, we have taken both $x$- and $y$-uncertainty into account with the orthogonal distance regression method \citep{odr89}.}
We also fit a Gaussian ring to the deprojected \Qr\ surface density map, and the ring is centered at  $r_{\rm center} = 45.6\pm0.2$ au with a $24.9\pm0.4$ au full width at half maximum (FWHM). We do not report the results from STIS  {or NICMOS}, since they have low spatial sampling and {are noisy near the corresponding IWAs}.

We measure the geometric parameters for the halo from the STIS data because they cover the largest field of view.  We measure a surface brightness power law index of $\gamma_{\rm out} = -2.68\pm0.04$. The power law indices measured at different position angles are also consistent with the integrated measurement within $1\sigma$. Therefore, we report the integrated radial profiles to reduce systematic uncertainty.\\

In the above calculation, we have assumed a flat disk (i.e., $h \ll 1$); however, the disk is not perfectly flat, and the line of sight passes through different radii at different heights. Under this scenario, for a non-flared disk,  we calculate that the radial separation will be modified by  multiplicative factors of \begin{equation}
(1\pm0.5h\tan\theta_{\rm inc})^{-1}.
\end{equation} For the HD~191089 system, these line-of-sight intersections {increase the uncertainties by} ${\sim}3\%$  {for $h=0.04$}. Therefore, the approximation of a flat disk will not bias the results for a disk with small scale height. {For a continuous vertical distribution of the dust that is following a Gaussian decay, this effect is then an upper limit since most of the scatterers are close to the mid-plane.} We thus ignore this effect given its {model-dependent} minor impact on the uncertainties.\\

We find that the radial power law indices, $\gamma_{\rm out}$, for the ring and the halo differ by ${>}10\sigma$. {Despite the \textit{physical} connection between the ring and the halo}, although the two power law indices are measured at different wavelengths, when the radial distribution of the dust is wavelength-independent, the indices are indicating that the halo is not a \textit{geometrical} extension of the ring's outer part.

\subsection{Scattering Phase Function (SPF)}\label{sec:spf}

\begin{figure}[htb!]
\center
\includegraphics[width=0.49\textwidth]{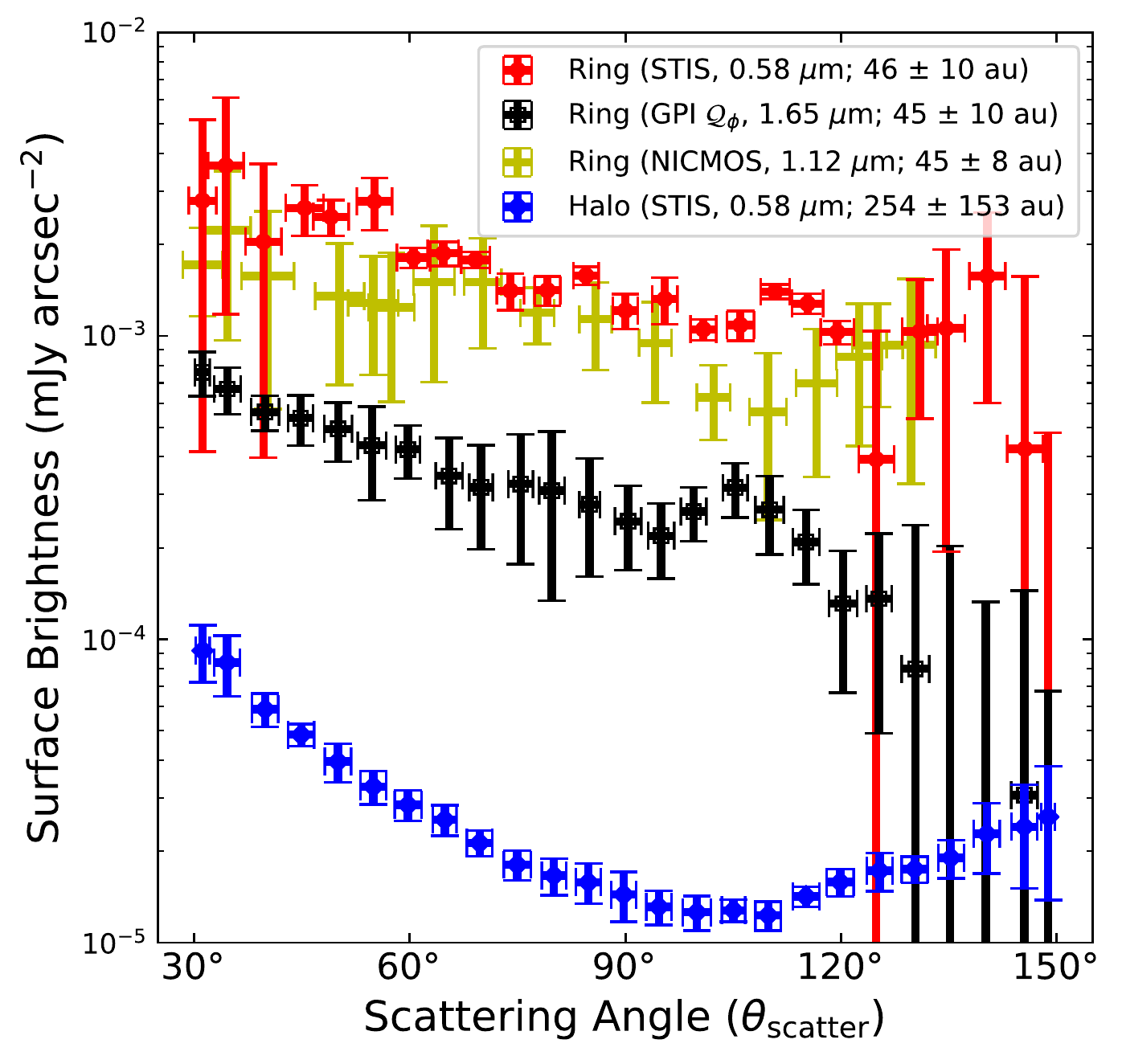}
\caption{SPFs for STIS-cRDI and NICMOS-NMF data, and the polarized phase function for the GPI \Qr\ data. The SPF for the halo of the STIS data is averaged from multiple SPFs at different stellocentric separations, thus minimizing the illumination and radial profile effects. See Figure~\ref{fig:stis-spf} for a linear-scale plot. Note: the radial extent of the ring is defined by the GPI \Qr\ image, and that of the other instruments is different due to pixel size.}
\label{fig:stis-spf-original}
\end{figure}

\citet[][]{hughes18} summarized the SPFs for different systems including zodiacal dust and debris disks, and found a tentative universal SPF trend for the dust in debris disks. To further investigate the similarities and differences of SPFs in different systems, we first derive the uncertainty for the scattering angles and the SPFs in \ref{appendix:spf}, then measure the empirical SPFs for the ring and the halo for our HD~191089 observations. 

{For the ring and the halo,} we present Figure~\ref{fig:stis-spf-original} the SPFs averaged for both sides {(i.e., the NE and SW sides) using STIS-cRDI and NICMOS-NMF results to minimize over-subtraction}. We observe different trends of the SPFs between the ring and the halo: in the STIS data, the halo is more forward and backward scattering than the ring; in the halo, the backward scattering is less strong than the forward trend. In addition, for the phase functions of the ring, the GPI \Qr\ polarized light image and STIS and NICMOS total intensity images have similar trends.

\begin{figure}[htb!]
\center
\includegraphics[width=0.49\textwidth]{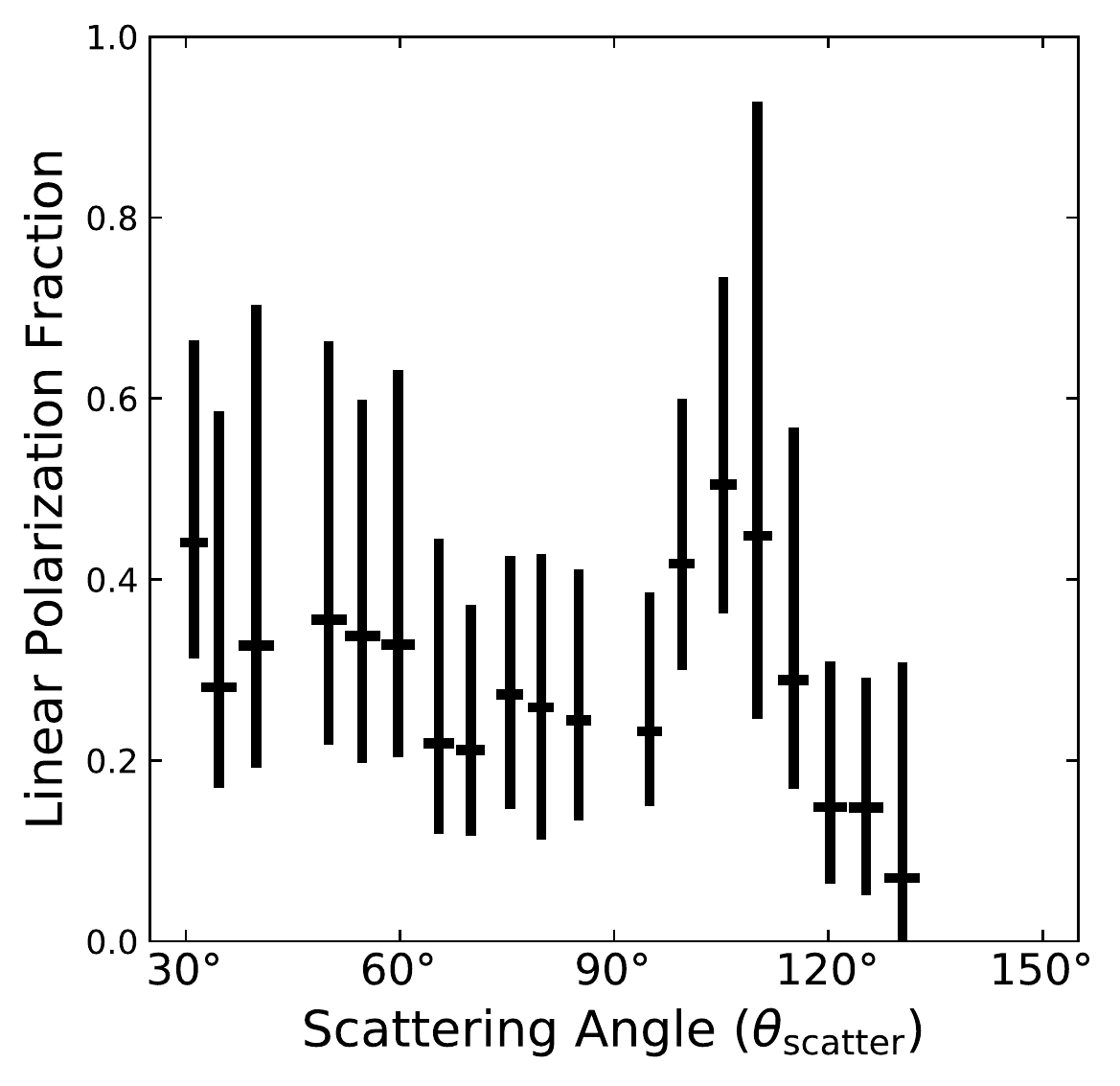}
\caption{ {Polarziation fraction for the ring as a function of scattering angle. The data are extracted from the ratio between the GPI \Qr\ (${\sim}1.65\,  \mu$m) and NICMOS-NMF (${\sim}1.12\, \mu$m) surface brightness profiles in Figure~\ref{fig:stis-spf-original}. We do not observe a clear trend of polarization fraction. However, it is possible that the polarization fraction peaks at ${\sim}110^\circ$.}}
\label{fig:pol-frac}
\end{figure}

{We measure the polarization fraction for the ring using the GPI \Qr\ and NICMOS-NMF images. Given the fact that we cannot recover the ring in total intensity with GPI $H$-band observations at $1.65\ \micron$ (Section~\ref{sec-obs-compare}), we instead use the $1.12\ \micron$ NICMOS-NMF observation that is both close to the GPI wavelengths and has less over-subtraction effect.} The polarization fractions derived from GPI \Qr\ and NICMOS-NMF images are around {$20\%$--$40\%$} with no clear trend  {(Figure~\ref{fig:pol-frac})}. {The polarization fraction is possibly peaking at ${\sim}110^\circ$, however it cannot be as firmly established as in previous measurements \citep[e.g.,][]{perrin14, milli19, frattin19}.} To better constrain the polarized fraction values, we need $H$-band total intensity observations to rule out the wavelength-dependent effect.

\begin{figure}[htb!]
\center
\includegraphics[width=0.49\textwidth]{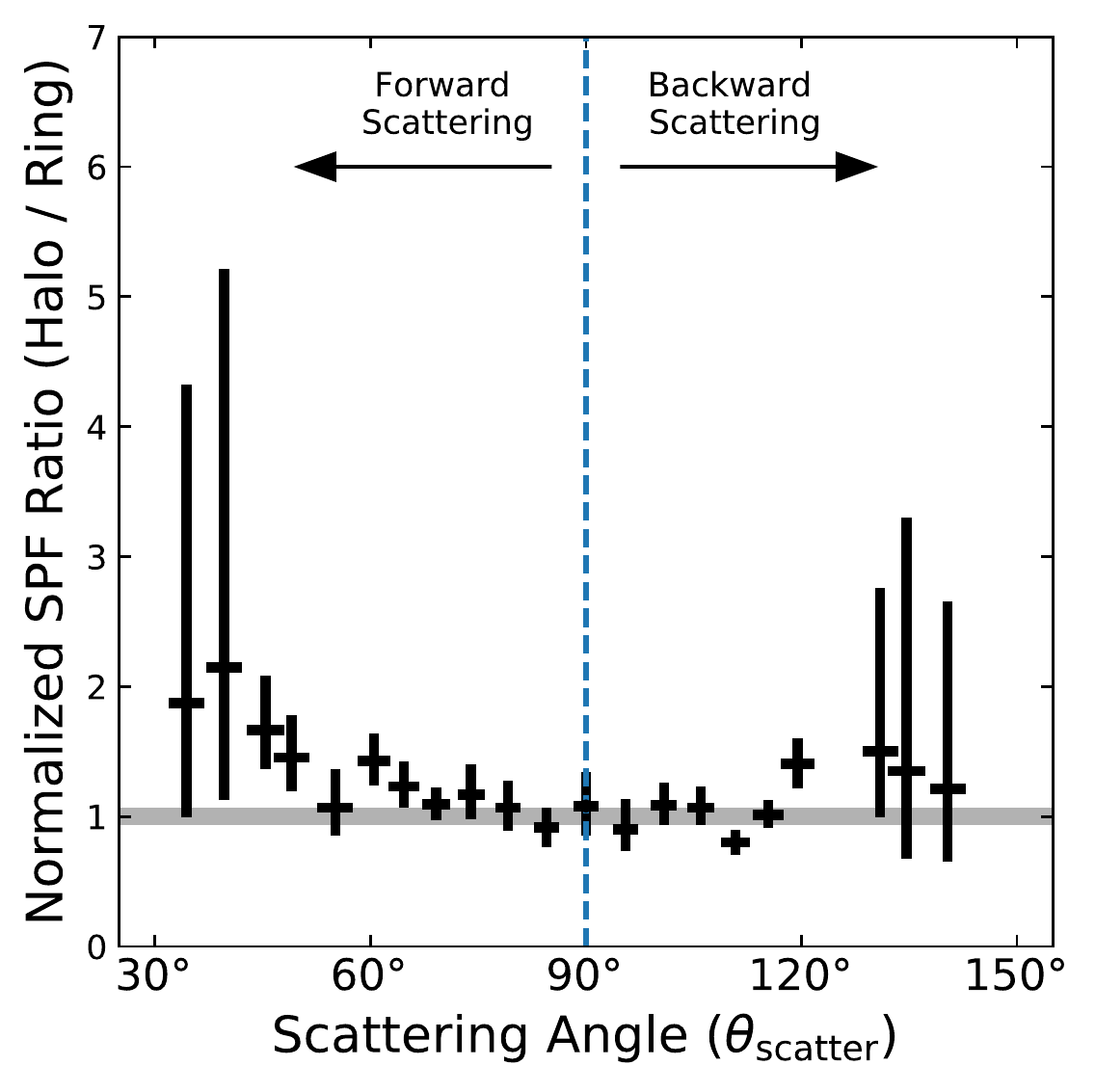}
\caption{Normalized SPF ratios for the STIS data. The ratios are  obtained by dividing the normalized SPF of the halo by that of the ring. The halo is likely both more forward scattering and backward scattering in the probed scattering angles.}
\label{fig:spf-halo-by-ring}
\end{figure}

To compare SPFs in the STIS image, we first normalize the SPFs by dividing their average surface brightness at $90^\circ\pm10^\circ$ scattering angle. We then divide the normalized halo SPF by that of the ring to illustrate the difference. In Figure~\ref{fig:spf-halo-by-ring}, the halo is likely both more forward scattering and backward scattering in the probed scattering angles.

\begin{figure*}[htb!]
\center
\includegraphics[width=0.9\textwidth]{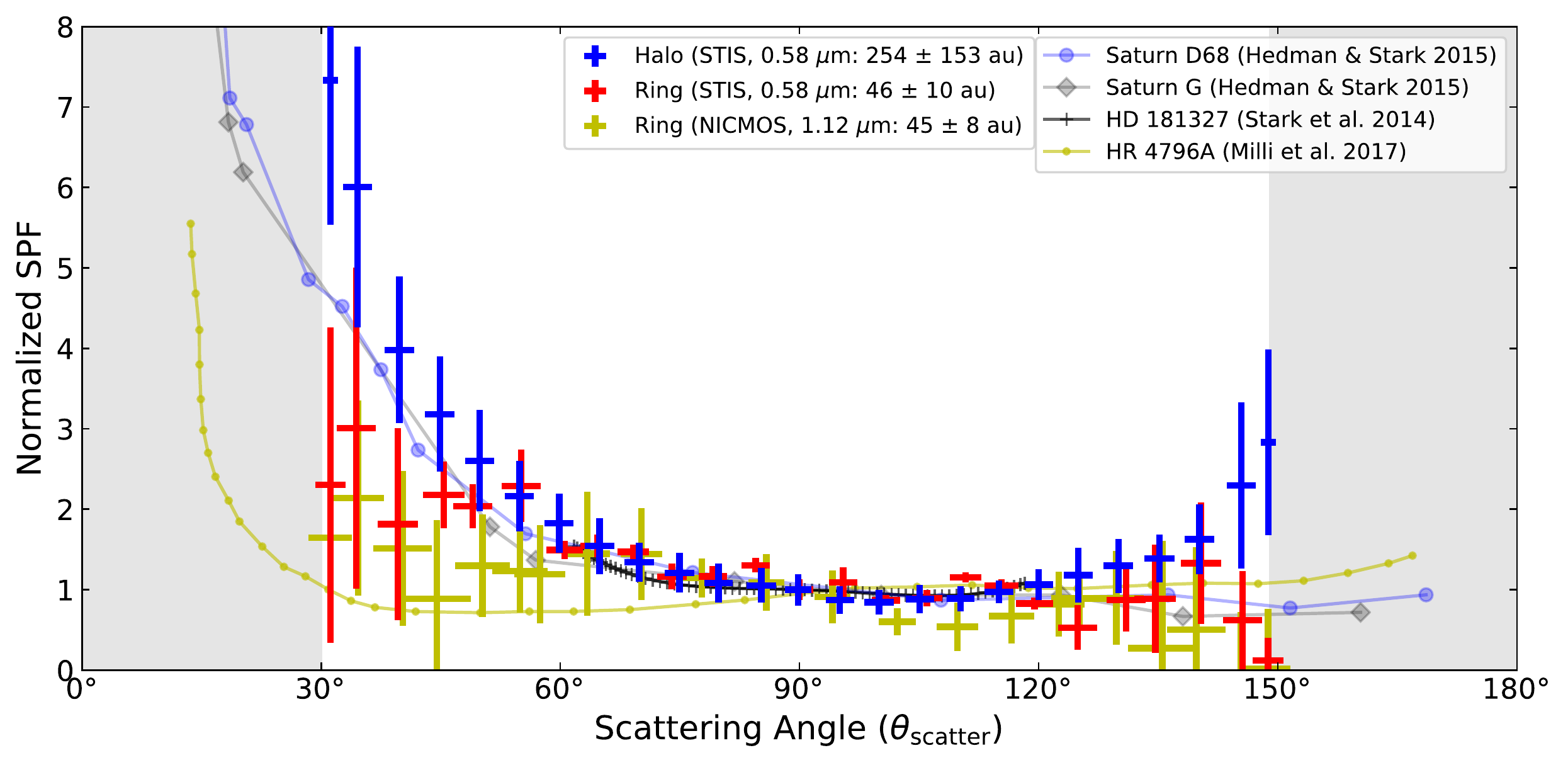}
\caption{Normalized SPFs for the STIS  and NICMOS total intensity data. The red and yellow error bars are the SPFs of the ring in the STIS-cRDI and NICMOS-NMF data, the blue ones are for the halo in the STIS data. From the SPFs, the ring and the halo are composed of two {\it distinct} populations of dust, with the halo dust more forward and backward scattering. For comparison, the observed SPFs for other systems are also plotted with lines.  {Note: due to the inclination of the HD~191089 system, the scattering angles in the shaded areas are not probed.}}
\label{fig:stis-spf}
\end{figure*}

 {To compare the HD~191089 SPFs with the ones in the literature}, we present the  {normalized} SPFs in linear scale for selected samples including both solar system objects (Saturn's D68 and G rings: \citealp{hedman15}) and circumstellar disk systems (HD~181327: \citealp{stark14}, HR~4796~A: \citealp{milli17}) in Figure \ref{fig:stis-spf}. Comparing with the previous studies, the HD~191089 ring SPF lies between the Saturn rings and the other samples, while its halo SPF lies above the Saturn rings.

\subsection{Disk Color}\label{sec:disk-color}
\begin{figure}[htb!]
\center
\includegraphics[width=0.45\textwidth]{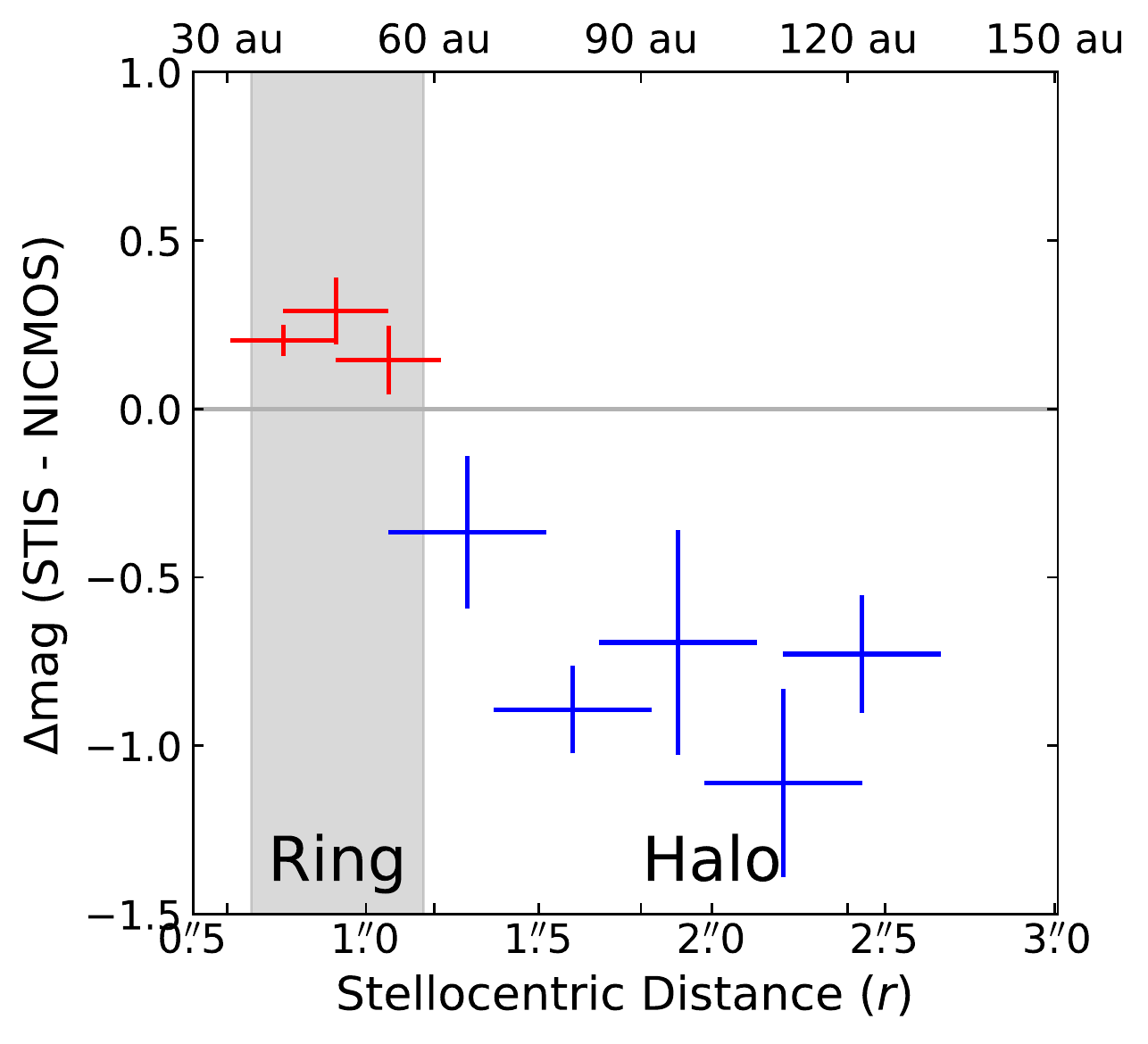}
\caption{Average magnitude difference relative to the star along the ansae between the $0.58\ \mu$m STIS-NMF and the $1.12\ \mu$m NICMOS-NMF observations. The ring (shaded area: the FWHM of the Gaussian ring in Table~\ref{tab:disk-measure}) scatters more flux in the longer NICMOS wavelengths, indicating red color of the dust. The halo has opposite trend, and instead scatters more flux in the shorter STIS wavelengths. The trend of the ratio is consistent with the ring containing larger dust than the halo.}

\label{fig:disk-color}
\end{figure}

{Using the \textit{HST} disk images, we calculate the color of the dust as follows: we first bin the $1.12\, \mu$m NICMOS-NMF image and the $0.58\, \mu$m STIS-NMF image to two images with pixel size of ${\sim}150$ mas to reduce correlated noise. We then divide the binned images by the corresponding {\tt pysynphot}  \citep{pysynphot} NICMOS or STIS counts of a $T=6450$ K blackbody to obtain the magnitude per pixel. We calculate the difference of the two magnitude maps, and present the radial profile along the ansae of the system in Figure~\ref{fig:disk-color}.}

For the ring, the dust scatters ${\sim}25\%$ more light ($\Delta$mag $\approx 0.25$) within the NICMOS F110W passband than that in STIS 50CCD, showing a red scattering property. For the halo, $\Delta$mag $\approx -1$. {Dust in the halo is expected to be generated in the ring through collisional cascade, then the smaller dust that is more sensitive to radiation pressure migrates outwards to form the halo \citep[e.g.,][]{strubbe06, thebault08}.} Assuming that scattered light images primarily probe the cross sections of the dust whose sizes are comparable to the observing wavelength, this red-to-blue trend from the ring to the halo in Figure~\ref{fig:disk-color} is consistent with this scenario.

\section{Disk Modeling}\label{sec-diskrt}
\subsection{Radiative Transfer Modeling Tool}\label{sec-rtmt}
We model the HD~191089 ring with the {\tt MCFOST} \citep{pinte06, pinte09} radiative transfer modeling code, and describe the dust using a Distribution of Hollow Spheres \citep[DHS:][]{min03, min05, min07}. As a derivative of the Mie theory where the dust grains are assumed to be spherical \citep{mie1908}, DHS adopts vacuum centers for these spherical dust grains.  To approximate small irregularly shaped dust, {the only additional parameter in DHS from Mie}---maximum vacuum fraction ($f_{\rm max}$)---parameterizes the central vacuum fraction in the dust that is uniformly ranging from 0 to $f_{\rm max}$.  DHS has been used to successfully reproduce the scattering phase function of linearly polarized scattered light with incident unpolarized light for quartz particles in laboratory \citep{min05}, characterize the spectral features of the interstellar medium \citep[e.g., ][]{min07, poteet15}, and better fit the scattering properties of the HR~4796A circumstellar disk system \citep{milli15}.

In our study, we adopt the geometrical parameters derived in Section~\ref{sec-describe} for the ring, and perform radiative transfer modeling with MCFOST using the DHS theory to probe the dust properties (e.g., dust mass, minimum dust size, composition, porosity, maximum void fraction.).  DHS is a computationally intensive technique due to the complex nature of the constituent dust, we thus adopt parallel computation in the Python environment and use the  {\tt DebrisDiskFM} package \citep{debrisdiskFM}\footnote{\url{https://github.com/seawander/DebrisDiskFM}}.  {The framework is} developed to efficiently explore debris disk properties through radiative transfer modeling.

 {{\tt DebrisDiskFM} is based on two software codes. We use {\tt MCFOST} (Version 3.0.33) to generate disk model images using given input parameters. We use {\tt emcee} \citep[Version 3.0rc1,][]{emcee}, which makes use of a Markov Chain Monte Carlo (MCMC) strategy with affine invariant ensemble samplers \citep{goodman10}, to obtain the posterior distributions for these parameters. With the two software codes, we use {\tt DebrisDiskFM} to distribute posterior calculations among multiple computation nodes in a computer cluster\footnote{We performed the calculations on the Maryland Advanced Research Computing Center (MARCC): \url{https://www.marcc.jhu.edu}.}, with each node calculating its own {\tt MCFOST} models in parallel. Specifically, for each combination of input parameters, we store the model in a unique folder, with the folder named by the hashed string for the array of the input parameters. We also append the folder name by a hashed random number, preventing multiple nodes simultaneously accessing the same folder and causing errors.}

\subsection{Modeling the Ring}\label{sec-disk-modeling-steps}

{We model the STIS-cRDI and GPI \Qr\ rings since they cover the largest wavelength range and have higher data quality.} To study the dust properties, we assume the dust is made of different types of grains, each with a pure composition. We adopt the three compositions in \citet{esposito18}: the amorphous silicate dust (i.e., ``astronomical silicates'', \citealp{draine84}, denoted by ``Si''), amorphous carbonaceous dust \citep[][denoted by ``C'']{rouleau91}, and ${\rm H_2O}$-dominated ice described in \citet{li98} to model the $\beta$ Pic disk (denoted by ``ice''). The size of the dust, $a$, follows a power-law distribution with index $q$, i.e.,
\begin{equation}
{\rm d}N(a)\propto a^{-q}\, {\rm d}a.
\end{equation}
The distribution is truncated at a minimum size of $a_{\rm min}$, and we set a maximum limit of the dust size, $a_{\rm max} = 1000$ $\mu$m. For the HD~191089 system, we set $q=3.5$ for the expected dust-size distribution for debris disks undergoing collisional cascade (e.g., theory and simulation: \citealp{dohnanyi69, pan12}; observation: \citealp{macgregor16, esposito18}). 

 {In this paper, we probe the following seven parameters of interest through disk modeling:
\begin{itemize}
\item Disk mass, $M_{\rm disk}$, which generally controls the overall brightness of the disk at different wavelengths;
\item Porosity;
\item Mass fraction for ``astronomical silicates'', $f({\rm Si})$;
\item Mass fraction for amorphous carbonaceous dust, $f({\rm C})$;
\item Mass fraction for ice, $f({\rm ice})$;
\item Minimum dust size, $a_{\rm min}$;
\item Maximum void fraction for DHS, $f_{\rm max}$.
\end{itemize}}

 {Given the fact that the composition parameters are interconnected, i.e., $f({\rm Si})+f({\rm C})+f({\rm ice})=1$, we only explicitly sample $f({\rm Si})$ and $f({\rm C})$. We also set the lower limit for $a_{\rm min}$ to be $0.5\,\mu$m based on the calculation of the blowout sizes for different dust compositions by \citet{arnold19}.} In the implementation of MCMC modeling of the system using the {\tt DebrisDiskFM} framework, for a given set of parameters, we first generate two parameter files for {\tt MCFOST} to represent the spatial sampling and field of view for the three instruments.  We then perform radiative transfer modeling with {\tt MCFOST}, using the DHS theory, to calculate the images for the three instruments at their central wavelengths in Table~\ref{tab:log}.

To simulate instrument responses, we convolve the disk models with {\tt TinyTim} PSFs \citep{krist11}\footnote{\url{http://www.stsci.edu/software/tinytim/}} for STIS, and with a 2-dimensional Gaussian profile for GPI (FWHM $=53.8$ mas, corresponding to 3.8 times the pixel scale for GPI {to match the GPI PSF}: \citealp{esposito18}).

 We compare the STIS model directly with the STIS-cRDI image. We compare the GPI model with the GPI \Qr\ image by first converting the Stokes $Q$ and $U$ models to a \Qr\ model, {then compare the PSF-convolved \Qr\ model with the observation}. With the models and observations, we maximize the following log likelihood function:
\begin{align*}\label{eq-loglike}
\log\mathcal{L}\left(\theta\mid X_{\rm obs}\right) = &-\frac{1}{2}\sum_{i=1}^{N}\left(\frac{X_{{\rm obs}, i} - X_{{\rm model}, i}}{\sigma_{{\rm obs}, i}}\right)^2\\
	&- \sum_{i=1}^{N}\log\sigma_{{\rm obs}, i} - \frac{N}{2} \log(2\pi). \numberthis
\end{align*}
In the above equation, $X$ is a flattened image with $N$ pixels. Subscripts $_{\rm obs}$ and $_{\rm model}$ denote observation and model, respectively. $\sigma_{{\rm obs}, i}$ is the uncertainty for $X_{{\rm obs},i}$ at the $i$-th pixel. {We only focus on the disk region (i.e., between $r_{\rm in}$ and $r_{\rm out}$ determined by GPI \Qr\ image) to minimize the influence from the halo.} In this paper, we assume the pixel noise follows a Gaussian distribution, and that all pixels are independent of each other (see, e.g., \citealp{wolff17}, for a proper treatment of correlated pixels).

\begin{deluxetable}{l|lcc}[htb!]
\tablecaption{{Independent} Dust Properties Retrieved from Ring Image Modeling\label{tab-mcfost-results}}
\tablehead{
Parameter\tablenotemark{a}	& Prior\tablenotemark{b}	& \multicolumn{2}{c}{Posterior\tablenotemark{c}}\\
Image & 		& GPI \Qr	& STIS }
\startdata
$\log_{\rm 10} M_{\rm disk}$ ($M_{\odot}$)	& $(-12, -4)$	&	$-7.462_{-0.010}^{+0.015}$	&  $-6.79_{-0.04}^{+0.03}$ \\
porosity				& (0, 1)						& 	$0.600_{-0.008}^{+0.012}$	& $0.01_{-0.01}^{+0.01}$ \\
$f({\rm Si})$		& (0, 1)			&	$0.005_{-0.005}^{+0.016}$		& $0.50_{-0.05}^{+0.04}$ \\
$f({\rm C})$		& (0, 1)	 		&	$0.004_{-0.004}^{+0.011}$		& $0.50_{-0.04}^{+0.05}$ \\
$f({\rm ice})$\tablenotemark{d}			& 		&	$0.989_{-0.010}^{+0.009}$	& $0.003$\tablenotemark{e}	\\
$\log_{\rm 10} a_{\rm min}$ ($\mu$m)	& $(-0.3, 2)$	 	&	$-0.115_{-0.007}^{+0.013}$	& $0.24_{-0.03}^{+0.01}$ \\
$f_{\rm max}$			& (0, 1)			&	$0.213_{-0.014}^{+0.026}$		& $0.001$\tablenotemark{e} \\ \hline
$\chi^2_\nu$ (GPI best fit)	&		&	1.05	&  240 \\
$\chi^2_\nu$ (STIS best fit) &		&	12	&	10	\\ \hline
$q$ 					& 	\multicolumn{3}{c}{3.5\tablenotemark{f}}		\\
$\log_{10}a_{\rm max}$ ($\mu$m)	& 	\multicolumn{3}{c}{3\tablenotemark{f}}		\\
Iteration Number	&\multicolumn{3}{c}{ {$9000$}\tablenotemark{f}} 			\\		
Burn-in			&\multicolumn{3}{c}{$2000$\tablenotemark{f}}     
\enddata
\tablenotetext{a}{The morphological parameters in Table~\ref{tab:disk-measure} are adopted.}
\tablenotetext{b}{The parameters are limited to 3 decimal digits, with uniform sampling in the prior range. For $M_{\rm disk}$ and $a_{\rm min}$, they are log-uniformly sampled.}
\tablenotetext{c}{16th, 50th, and 84th percentiles.}
\tablenotetext{d}{The mass fraction for ice is $f({\rm ice})=1 - f({\rm Si})-f({\rm C})$, thus only Si and C are explicitly sampled.}
\tablenotetext{e}{$95$th percentile.}
\tablenotetext{f}{Kept fixed.}
\end{deluxetable}

\subsubsection{GPI \Qr\ Image}\label{sec-mcmc-gpi}
\paragraph{One-Step GPI}
While the ring is resolved with all three instruments, we first only fit the GPI \Qr\ image to investigate the dust properties. Using the flat priors presented in Table~\ref{tab-mcfost-results}, we assign 60 chains to explore the parameter space, and run the MCMC modeling procedure for $9000$ steps. We discard the first $2 000$ steps that are identified as burn in stage, and calculate the posterior distributions using the last $7000$ steps (a total of $4.2\times10^5$ models). 

Table~\ref{tab-mcfost-results} reports the credible intervals from the posterior distributions in Figure~\ref{fig:posterior-gpi} (generated by {\tt corner}: \citealp{corner}). {We notice the small uncertainties in the results, these uncertainties are underestimated since the correlated noise is ignored in our likelihood function \citep[e.g.,][]{wolff17}. In addition, we also argue that these small uncertainties are also the results from the limited dust models, that various scenarios may lead to small uncertainties (e.g., spatial distribution asymmetries, non--power law surface density distribution), thus the inclusion of correlated noise may still not lead to physical interpretations of the retrieved parameters.}

\subsubsection{STIS Image}
\paragraph{One-Step STIS}
We fit the STIS image using the same priors as for the GPI image. We run MCMC modeling for $9 000$ steps with $60$ chains, and discard the first $2000$ as burn-in steps as for the GPI \Qr\ image. For this approach, we also present the posterior results in Table~\ref{tab-mcfost-results} and Figure~\ref{fig:posterior-gpi}.

\paragraph{Two-Step MCMC}
Noticing the two distinct sets of MCMC posteriors for the two images, we try to establish the connection between the GPI image and STIS image with a two-step MCMC fitting: we obtain the posterior distributions for the seven variables from the GPI \Qr\ image, then use them as the priors to fit the STIS image. In this way, we use the GPI posterior distribution as the null hypothesis, and test it on the STIS image. 

We use Probability Integral Transform (PIT: \ref{appendix:2smc}) to draw samples from the GPI posterior distributions. To find the posterior ranges from the STIS image, we only explore the GPI posteriors between their 2.5th and 97.5th percentiles as an analogy to the conventional level of significance for one parameter (i.e., $p$-value $\geq0.05$ for double-tailed  distribution). We run MCMC modeling for $3500$ steps with $60$ chains, and discard the first $1000$ as burn in stage. 

For the STIS modeling with the PIT approach, we generate the posterior distributions and overplot them on the one-step GPI data in Figure~\ref{fig:posterior-pit} (Appendix~\ref{sec:pit-gpi-stis}). From the following, we notice the statistical deviation of the STIS posteriors from the GPI posteriors: (1) the STIS posteriors using PIT are adjacent to the 2.5th- or 97.5th-percentile boundaries of the GPI posteriors, indicating the trend of drifting away from the null hypothesis; and (2) the STIS posteriors using PIT have extremely narrow credible intervals, indicating there is no statistically preferred solution within the explored intervals.

\subsubsection{Implications}\label{sec-implications}
We present the results from independent GPI and STIS modelings, and discuss the implications as follows.

\paragraph{SED}

\begin{figure}[htb!]
\center
\includegraphics[width=0.47\textwidth]{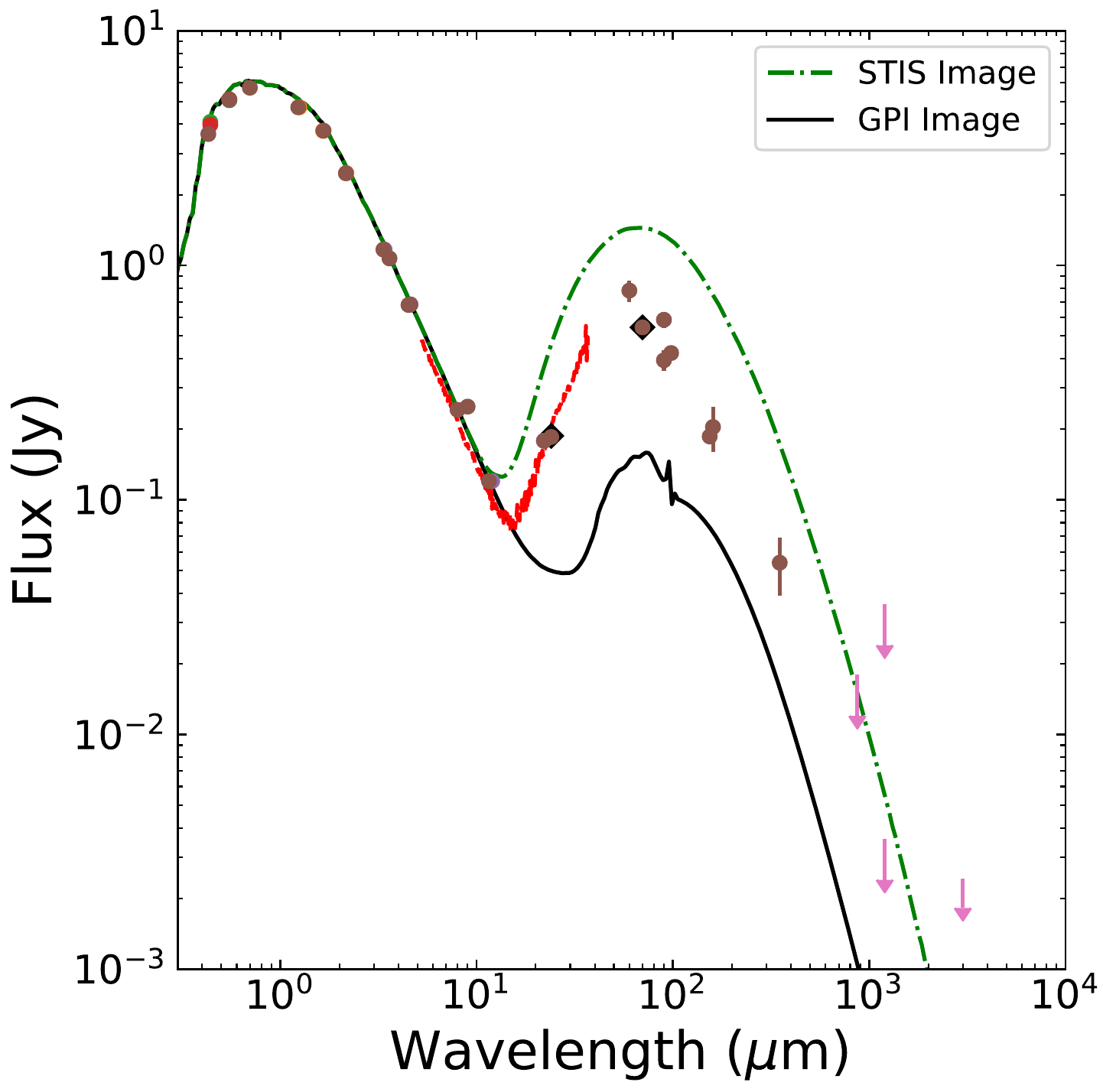}
\caption{ {Observed SED and SED models for HD~191089. Green dash-dotted line: SED generated using the best-fit STIS parameters. Black line: SED generated using the best-fit GPI \Qr\ parameters. Other: observation data obtained from \citet{soummer14}.}}
\label{fig:obs-model-sed}
\end{figure}

We generate the SEDs corresponding to the best fit parameters for the STIS and GPI \Qr\ images with {\tt MCFOST}, and compare them with the observed one in Figure~\ref{fig:obs-model-sed}. We notice that the SED from the STIS best fit over-predicts the emission from the system, while the SED from the GPI \Qr\ best fit under-predicts the emission.  {Even though different sizes of dust dominate the SED and the scattered light images, the inability to reproduce the observed SED adds another evidence that our study using DHS models cannot provide a consistent model satisfying all observables. For instance, the true dust albedo differs from the one predicted in DHS models.}

\paragraph{Mie vs DHS} Although we do not obtain a set of parameters that is able to explain both datasets, we can still focus  on the $f_{\rm max}$ parameter (i.e., maximum void fraction for the DHS theory). {We argue that it is the most profound parameter that is added to the DHS theory from the Mie theory---the $f_{\rm max}$ parameter approximates dust grains from spheres to aggregate-like structures \citep{min03, min05, min07}, and we believe it is one of the keys to understanding the scattered light properties of the dust.}

For the STIS image in total intensity, we obtain $f_{\rm max} \approx 0$. In DHS theory, this value corresponds to the Mie theory scenario. For the GPI \Qr\ image in polarized light, we obtain $f_{\rm max}\approx 0.21$. In this scenario, the DHS theory is preferred to the Mie theory, {and the $f_{\rm max}$ parameter is smaller than what is in the interstellar dust \citep[e.g., 0.7:][]{min07}, suggesting dust properties are different under various environments. Given that the two datasets differ in two fundamental way (different wavelengths, and total intensity/polarization observation), we cannot determine which is the root cause of this discrepancy. However, this discrepancy in the $f_{\rm max}$ parameter is still informative since it is an approximation of the dust structure---the discrepancy indicates that neither Mie nor DHS is able to well approximate the structure of the dust seen in scattered light images.}

\begin{figure*}[htb!]
\center
\includegraphics[width=\textwidth]{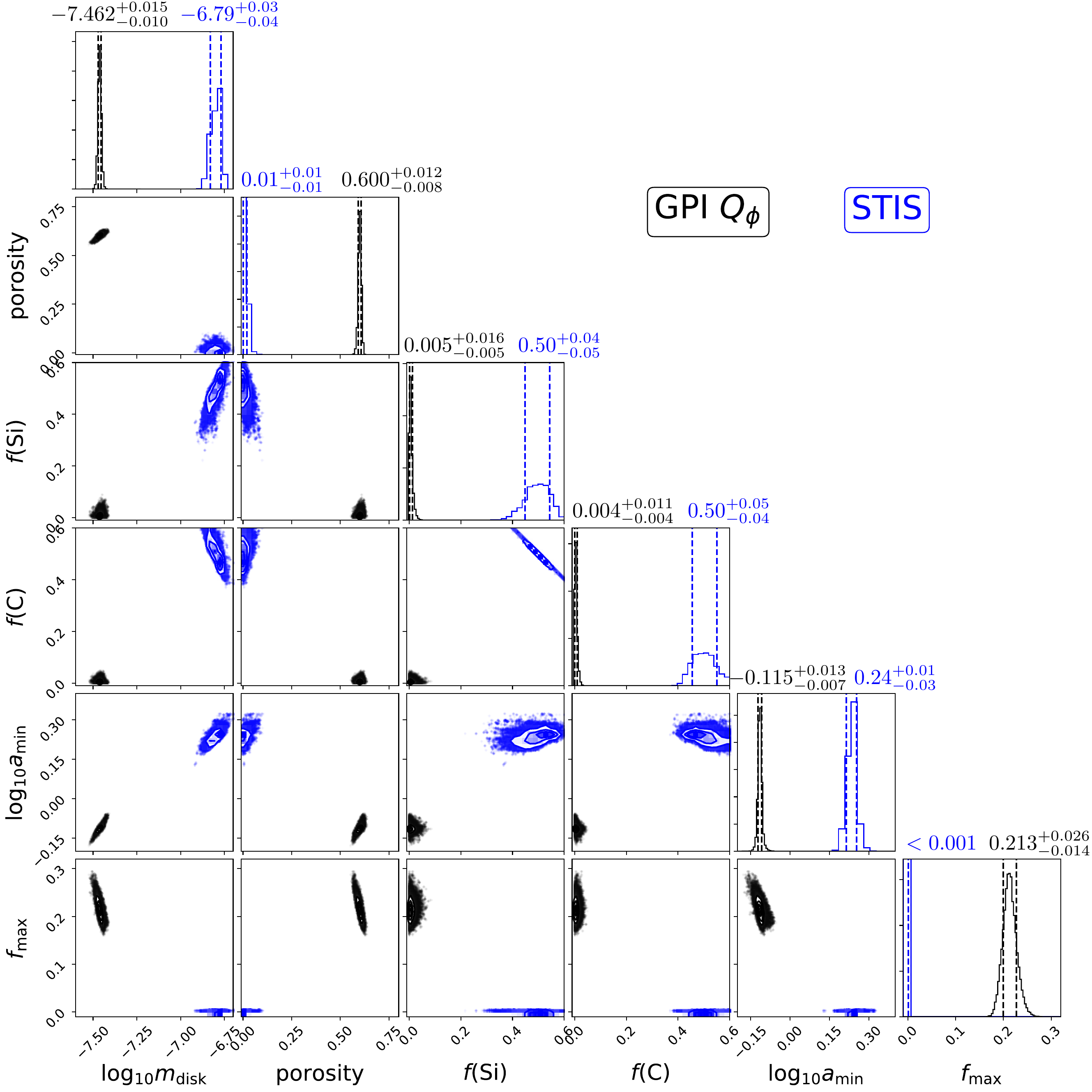}
\caption{Posterior distribution for the variables in Table~\ref{tab-mcfost-results} used in disk modeling for the GPI \Qr\ (black) and STIS (blue) images. The vertical dashed lines show the (16, 84)-th percentiles for the data. See Figure~\ref{fig:posterior-pit} for the posterior distribution focused on GPI modeling.}
\label{fig:posterior-gpi}
\end{figure*}

{\paragraph{Distinct Parameters}
In the radiative transfer modeling of the scattered light images, we retrieve distinct sets of parameters in Table~\ref{tab-mcfost-results} and Figure~\ref{fig:posterior-gpi}. Although the best fits are not statistically consistent with each other, we categorize the difference in the dust properties in the ring into two groups:
\begin{itemize}
\item Physically Possible: minimum dust size. The retrieved minimum dust size is ${\sim}0.8\,\mu$m for GPI \Qr and ${\sim}1.7\,\mu$m for STIS. Both values are within a factor of ${\sim}2$, and the inclusion of correlated noise in the likelihood function may resolve the discrepancy. The values are reasonably consistent with blowout size calculations, although we caution that these are themselves highly dependent on disk composition, porosity and aggregate structure---the blow out size for dust can vary by an order of magnitude for different composition and porosity (see the \citealp{arnold19} calculation for HD~181327, a star similar to HD~191089).
\item Physically Impossible: porosity and composition fraction. {Specifically, these parameters takes values in limited ranges (i.e., from $0\%$ to $100\%$), and the discrepancy is currently at ${\sim}50\%$ level. Even though the discrepancies for these parameters can be alleviated with larger uncertainties, their physical meaning are then uninformative---a possible solution is to increase the uncertainties to ${\sim}50\%$ (assuming the inclusion of correlated noise is able to achieve it), however that would render these parameters meaningless, since the large uncertainties would not reject any values in the physically plausible values (i.e., from $0\%$ to $100\%$). }
\end{itemize}
We note that the above parameter values are based only on MCMC fitting results, and neither may be correct given the limitation of DHS or Mie models. {Specifically, these models are optimized for spectral fitting \citep[e.g., ][]{min07, poteet15}, but neither model is able to properly constrain the composition from scattered light images \citep[e.g.,][]{milli19}. See Section~\ref{sec-dis-mcrt-miedhs} for more discussion on the dust properties retrieved from radiative transfer modeling.}
}

\subsubsection{Other Attempts}
In addition to the above modeling efforts, we have performed separate modeling attempts with loosen prior constraints.

\paragraph{For GPI} Set the lower limit in the prior for $a_{\rm min}$ to be 0.01 $\mu$m, and keep $q$ unconstrained. We observed a steeper $q\approx4.15$ with $a_{\rm min}\approx0.02\,\mu$m to describe the GPI \Qr\ image, but it still does not recover the STIS flux density. In addition, although these smaller dust is not blown out by radiation pressure, its collisional cascade suppliers (slightly larger dust) are blown out \citep[e.g.,][]{burns79, silsbee16, arnold19}, thus this scenario is not stable.

\paragraph{For GPI \& STIS} Simultaneously model the STIS and GPI images with the conditions above  {(i.e., as in the previous bullet point)}. However, the best fits indicate a bimodal distribution, with one better recovering the STIS image and the other better recovering the GPI image. In the former model, the GPI \Qr\ model displays negative polarization at the smallest scattering angles, failing to properly recover the observation; in the latter, the STIS model does not recover the ansae in the observed data.\\

\begin{figure*}[htb!]
\center
\includegraphics[width=\textwidth]{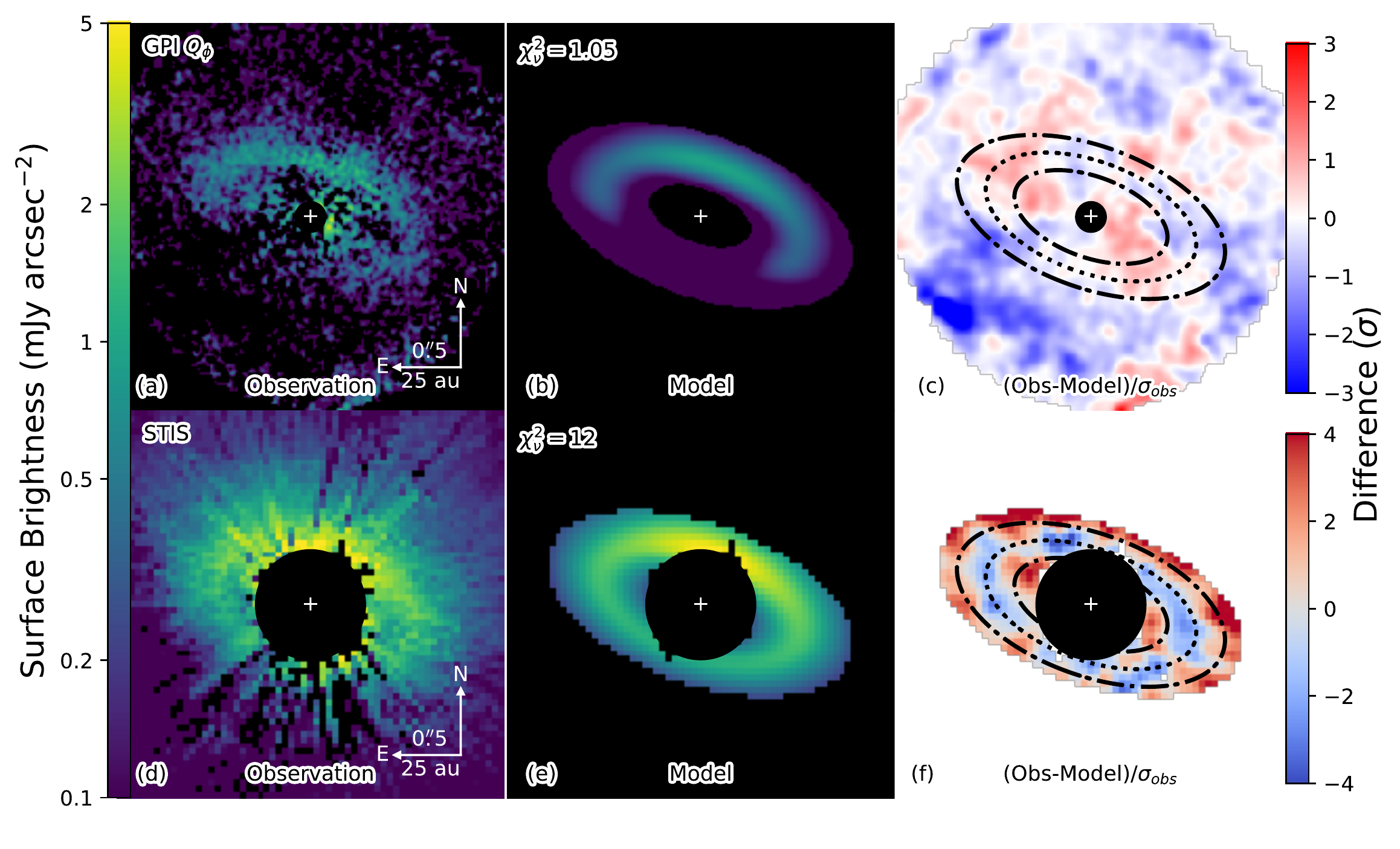}
\caption{Best fit results from MCMC fitting (performed individually). Observation ({\it left}), {\tt MCFOST} model ({\it middle}), and residual map ({\it right}) for the GPI data ({\it top}) and the STIS data ({\it bottom}). The reduced $\chi^2$ values are computed for the non-masked regions, the masked regions are denoted by black areas in the middle column. On the residual maps (both are smoothed with Gaussian kernels of $\sigma=50$ mas to remove high frequency noise), the dotted and dash-dotted ellipses represent the peak location and FWHM region of the  {GPI Gaussian} ring.}
\label{fig:obs-model-dev}
\end{figure*}

Based on our modeling efforts, we conclude that the STIS and GPI datasets cannot be consistently reproduced with a single model. As a result, we present our separate models for the GPI \Qr\ image and the STIS image in Figures~\ref{fig:posterior-gpi} and \ref{fig:obs-model-dev}. See Section~\ref{sec-dis-mcrt-miedhs} for more discussion on applying the DHS theory to disk modeling.

\section{Discussion}\label{sec-diss}
\subsection{Spatial Distribution}\label{sec-discuss-measure}

 {Although our radiative transfer modeling efforts cannot explain the scattering properties of the ring, we are confident in the results of our geometrical analysis, since they are based only on the surface brightness distribution of the system.}

\subsubsection{Ring Measurables}\label{sec-primaring-ring-measure}

{\it Ring Clearing Radii} ($r_{\rm in}$ and $r_{\rm out}$): \citet{churcher11} observed the ring at $18.3\micron$, reporting a dust belt from 26 to 84~au\footnote{Updated with the {\it Gaia} DR2 distance to HD~191089.}, which is consistent with our fitting results of $r_{\rm in} = 26\pm4$~au and $r_{\rm out} = 78\pm14$~au at $1\sigma$ level. The position angle of the major axis, as well as the inclination, is better constrained with our high spatial resolution data in scattered light with GPI.

{\it Brightness Asymmetry}: For the ring, we are not able to find brightness asymmetry beyond ${\sim}10\%$ or $1\sigma$ with the GPI and STIS data. Although a tentative ${\sim}20\%$ asymmetry was observed at the $1.8\sigma$ level in the $18.3\mu$m observation by \citet{churcher11}, if the dust follows the same spatial distribution at these wavelengths, the $18.3\ \micron$ emission asymmetry is likely resulted from statistical or instrumental fluctuation.
 
{\it Planet Perturber}: In the GPI $H$-band total intensity HD~191089 observations, we did not detect any point source (Figure~\ref{fig:SNR-methods}). We report $5\sigma$ point-source contrast limits of ${\sim}1\times10^{-5}$--${\sim}3\times10^{-6}$ between $0\farcs3$ and $0\farcs8$ with the forward modeling planet detection method to correct for self- and over-subtraction in \citet{ruffio17}. Using these contrast limits, if a planet is shepherding the ring, for a system with an age of {$22$} Myr and using the evolution tracks in \citet{spiegel12}, its mass is expected be smaller than ${\sim}5~M_{\rm Jupiter}$.

Using the \citet{morrison15} analysis for the outer edge of a planet's chaotic zone, and assuming this outer edge is the inner edge of the ring of a debris disk, the upper limit on planet mass can be translated to the lower limit on the semi-major axis of the planet's orbit:
\begin{equation}\label{eq-morrison15}
a_{\rm p} = \frac{r_{\rm in}}{1+1.7 (5~M_{\rm Jupiter}/M_{\star})^{0.31}}.
\end{equation}
When we substitute the measured $r_{\rm in}$ into the above equation, we obtain a lower limit of $a_{\rm p} = 20\pm3$~au for the semi-major axis of the planet.

\subsubsection{Ring as an Exo--Kuiper Belt}
\begin{figure*}[htb!]
\center
\includegraphics[width=.47\textwidth]{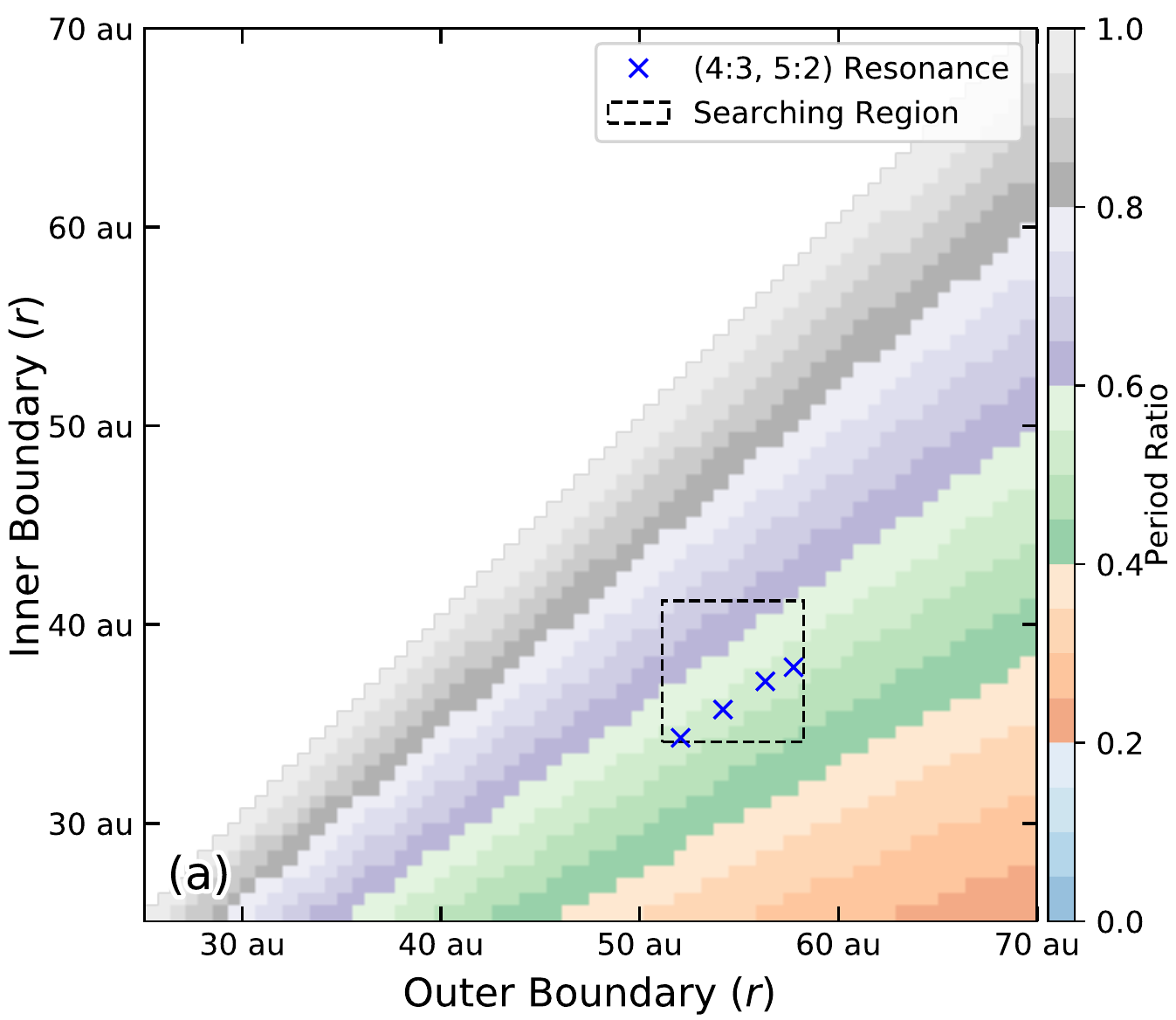}
\includegraphics[width=.47\textwidth]{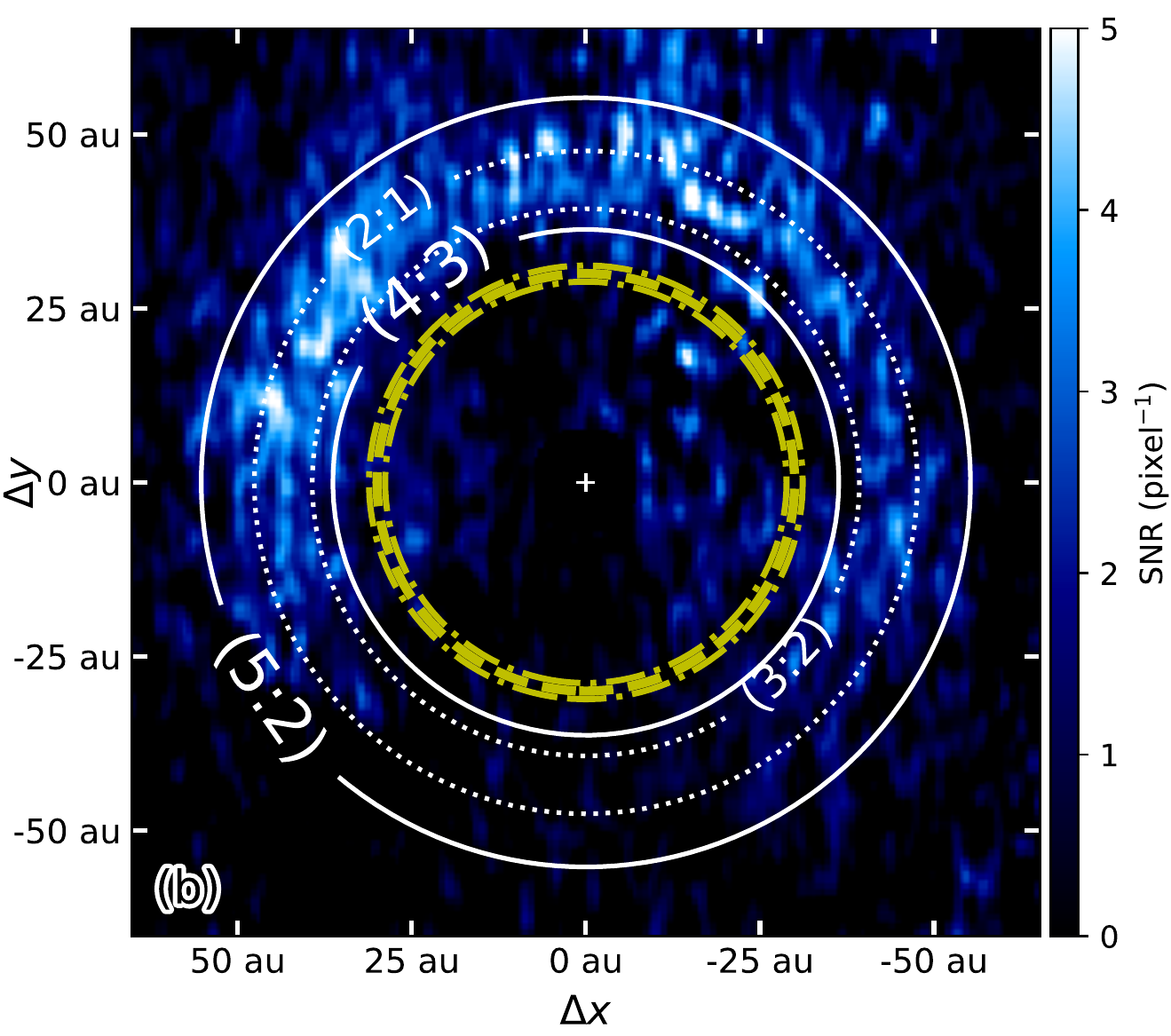}
\caption{({\bf a}): Searching for the mean-motion resonance in HD~191089's ring with the deprojected disk on the right, the colorbar shows the orbital period ratio between the inner and outer boundary. The searching range is marked with a dashed rectangle, and the (4:3, 5:2) mean-motion resonance radii for the inner boundary and outer boundary are marked with blue crosses. ({\bf b}): Sketch of the mean-motion resonance orbits between the deprojected HD~191089 ring and a hypothesized potential well ($r_{(1:1)}=29.9\pm1.2$~au, marked with yellow dashed and dash-dotted lines). If the extent of the primary disk matches the resonance orbits, the $4:3$ orbit is at $36.2\pm1.4$~au, with the $5:2$ orbit at $55\pm2$~au (white solid lines); the corresponding $3:2$ and $2:1$ orbits are marked with dotted lines.}
\label{fig:mean-motion-search}
\end{figure*}

The spatial extent of the ring ($r_{\rm center}=45.6\pm0.2$~au, FWHM $=24.9\pm0.4$~au for a Gaussian {ring}) resembles that of the Solar System's Kuiper Belt ($30$--$50$~au, \citealp{stern97, bannister18}). To investigate the scenario where the ring is an extrasolar version of the Kuiper Belt, which is perturbed by a corresponding planet (i.e., Neptune), we first deproject the SNR map of the GPI observation to a face-on view, then search for possible mean-motion resonance orbits for the inner and outer radii of the disk. Based on Kepler's third law, we search for the period ratios corresponding to different combinations of stellocentric radii in Figure~\ref{fig:mean-motion-search}, where the period ratios are mapped to simple fraction values that correspond with the strongest resonance orbits of a hypothetical potential well. 

We set the boundary ranges to be consistent with pixel-wise ${\rm SNR}\approx1$: among different combinations of the strongest resonances, we obtain 4 pairs of $4:3$ and $5:2$ resonance orbits to resemble the extent of the Solar System Neptunian resonance orbits in \citet{chiang03}. For the other resonance pairs, their boundary ranges are either too narrow to cover the GPI disk,  {or these pairs cannot be resolved because of the limitation of instrumental spatial resolution}, or these solutions are consistent with the ranges of the $4:3$ and $5:2$ pairs, {therefore} they are not presented or analyzed in this paper.

With the 4 resonance radii pairs, we are able to compute the location of the potential well (i.e., $1:1$ resonance) at $r_{(1:1)}=0\farcs60\pm0\farcs02$, correponding with a stellocentric distance of  $r_{(1:1)}=29.9\pm1.2$~au. The mean resonance radii, and the hypothesized $1:1$ orbit of the potential well, are shown in Figure~\ref{fig:mean-motion-search}. To confirm these resonances, deeper high resolution and high SNR observations are needed to firmly establish the edges of the ring. If the $1:1$ gravitational potential is caused by a planet, it is likely of small mass and requires the future {\it LUVOIR} or {\it HabEx} missions for observation.

\subsubsection{Halo: Radial Distribution}\label{sec-outer-halo-radial-vary}
\paragraph{Overall Distribution} In the STIS data, the halo extends to $r_{\rm out} = 640\pm130$~au, with a surface density power-law index of $\Gamma_{\rm out} = -0.68\pm0.04$. This power-law distribution index for the surface density profile is shallower than $-1.5$, i.e., the classical expectation for the halo of debris disks \citep[e.g.,][]{strubbe06, thebault08}. It is also shallower than the power law index of $-1$, which is expected for the steady state radial motion of the dust \citep{jewitt87}. Using the dust size--free approximation in \citet{jewitt87}, a gravity-dominated slow-down of the radial motion of the dust corresponds a power law index of $-0.5$, and a constant acceleration results into an index of $-1.5$. 

A power law index of $-0.68_{-0.04}^{+0.04}$ between the two values $(-0.5, -1.5)$ is thus caused by the joint effect from multiple force sources. In debris disks, the $-1.5$ power law index has already taken into account both the slow-down from gravity and the acceleration from radiation pressure; the $-0.68_{-0.04}^{+0.04}$ power law index is thus calling for additional slow-down sources, and the slow-down by interstellar medium is a plausible candidate. {In fact, a similar surface density profile has been observed in} the outskirts of the HR~4796~A halo, which has an index of $-0.7$  and has a large-scale structure that is strongly suggestive of interaction with the interstellar medium \citep{schneider18}.

Following the \citet{jewitt87} analytic derivation relating surface density to the radial speed of the dust, we assume the size distribution of the dust is independent of its stellocentric distance. Under this assumption, if the dust has an outward radial speed of $v(r)\propto r^x$, where $r$ is the radial separation, then the surface density will be $\Gamma(r)\propto r^{-x-1}$. In this way, the dust in the halo of HD~191089 has a radial speed of $v(r)\propto r^{-0.32\pm0.04}$.

\begin{figure}[htb!]
\center
\includegraphics[width=.47\textwidth]{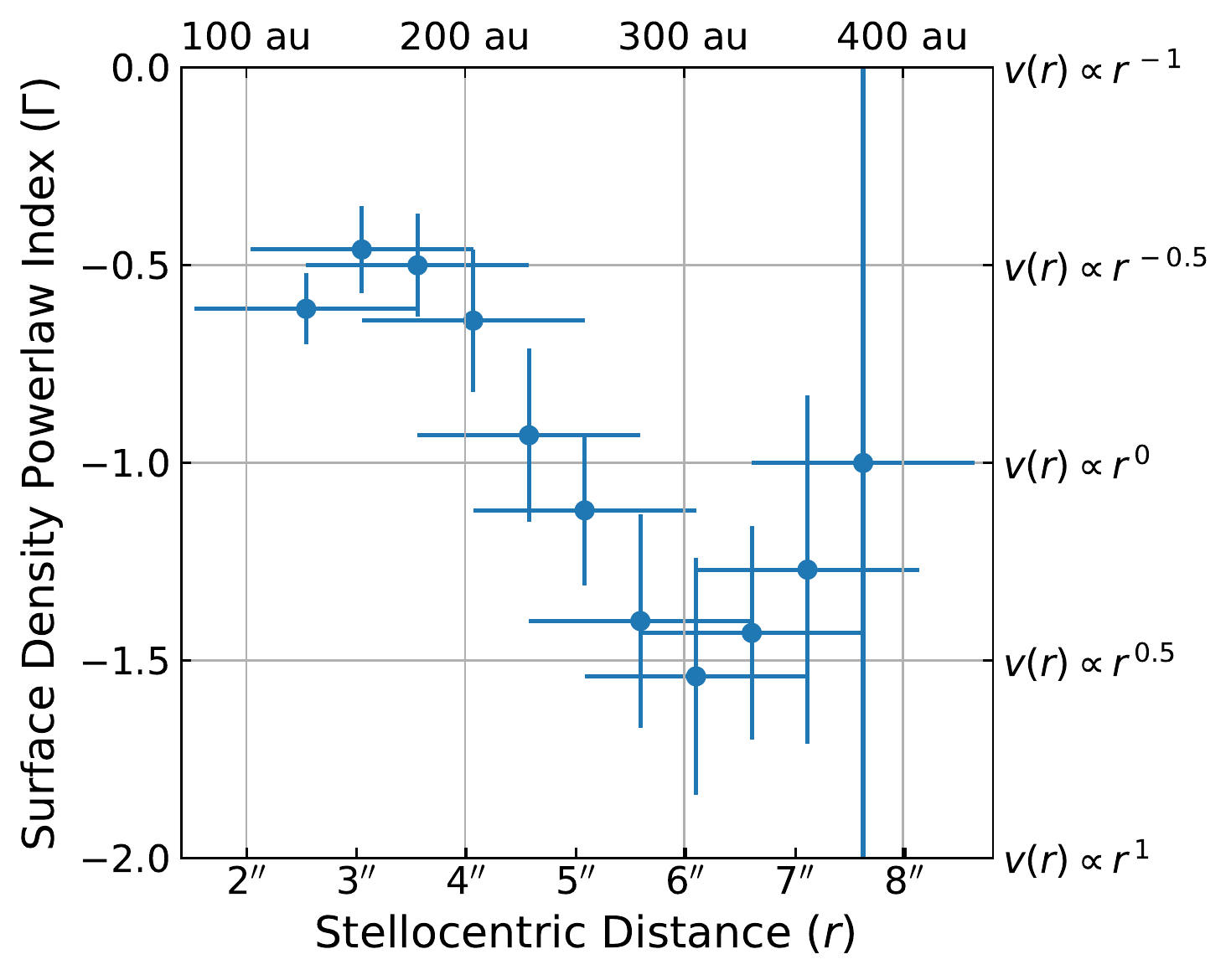}
\caption{Power law indices for the surface density radial profiles of the halo at different stellocentric distance, and the corresponding radial speed dependence using the derivation in \citet{jewitt87}. The horizontal error bars are the regions where the radial profiles are calculated, i.e., ${\pm}1\arcsec$.}
\label{fig:outer-powerlaw-trend}
\end{figure}

\paragraph{Local Distribution} {The surface density power law of the HD~191089 halo deviates from the classical model at ${>}10\sigma$ level, calling for detailed investigation for the local variation of the halo. As an attempt} to investigate the variation of the surface density distribution at different stellocentric distance, we compute the power law indices of the surface density radial profile for the halo at different radii in Figure~\ref{fig:outer-powerlaw-trend}. {Assuming the dust size distribution is independent of its stellocentric distance, we also present the \citet{jewitt87} analytical derivation between radial speed and surface density power law index under difference scenarios. At different stellocentric distances, we observe that
\begin{itemize}
\item  Interior to ${\sim}200$ au, the radial speed of dust decreases as stellocentric distance increases. This indicates the decrease of the net inward force that slows down the outward motion of the dust.
\item  Between ${\sim}200$ au and ${\sim}300$ au, the radial speed reaches a constant then increases as stellocentric distance increases. At ${\sim}300$ au, the surface density power law index reaches that for the classical model for debris disk halo \citep[e.g.,][]{strubbe06, thebault08}.
\item  Exterior to ${\sim}300$ au, albeit with large uncertainty, the radial speed marginally increases then reaches a constant as stellocentric distance increases.
\end{itemize}

Given the complex 2-dimensional residual structure in the single SPF--corrected distribution of the halo (Section~\ref{sec-spf-dev}), we do not further discuss the trend of the local distribution power law indices. In addition, our analysis is based on the assumption that dust size does not vary as a function of stellocentric separation, however the assumption is invalid for collision-dominated debris disks \citep[e.g.,][]{strubbe06, thebault08}. Therefore a full dynamical modeling of the halo is needed to better explain the local variations of the halo.}\\

\subsubsection{Halo: Surface Density Variation}\label{sec-spf-dev}
{In the classical model for debris disk halo, the dominant dust size for optical depth decreases as stellocentric distance increases \citep[e.g.,][]{strubbe06, thebault08}. Using the STIS observations of the HD~181327 halo, \citet{stark14} found a consistency between the observed SPF change (under Mie theory) and the classical model. However, we do not find a clear SPF variation trend for the HD~191089 halo. To investigate the HD~191089 halo, we adopt the averaged halo SPF from measurement, and explore the 2-dimensional surface density variation for the HD~191089 halo.}

To investigate the deviation of the scattering properties from a same SPF for the halo in STIS at different stellocentric radii, we  {first} scale the whole halo by the surface {brightness} radial distribution as measured for Figure~\ref{fig:radialprofile}, thus eliminating both the distance-dependent illumination and radial density distribution factors{. We then} divide the image by the empirically averaged SPF for the halo in Figure~\ref{fig:stis-spf} based on the scattering angles for each pixel. The scaled STIS image is then deprojected to a face-on view and rotated to align the major axis with the $x$-axis, and subtracted by the median to show the first-order deviation from an identical SPF in all of the halo. Based on the quality of the NICMOS-NMF data, only the $\gamma_{\rm out}=-2.68$ correction is applied.

\begin{figure}[htb!]
\center
\includegraphics[width=0.5\textwidth]{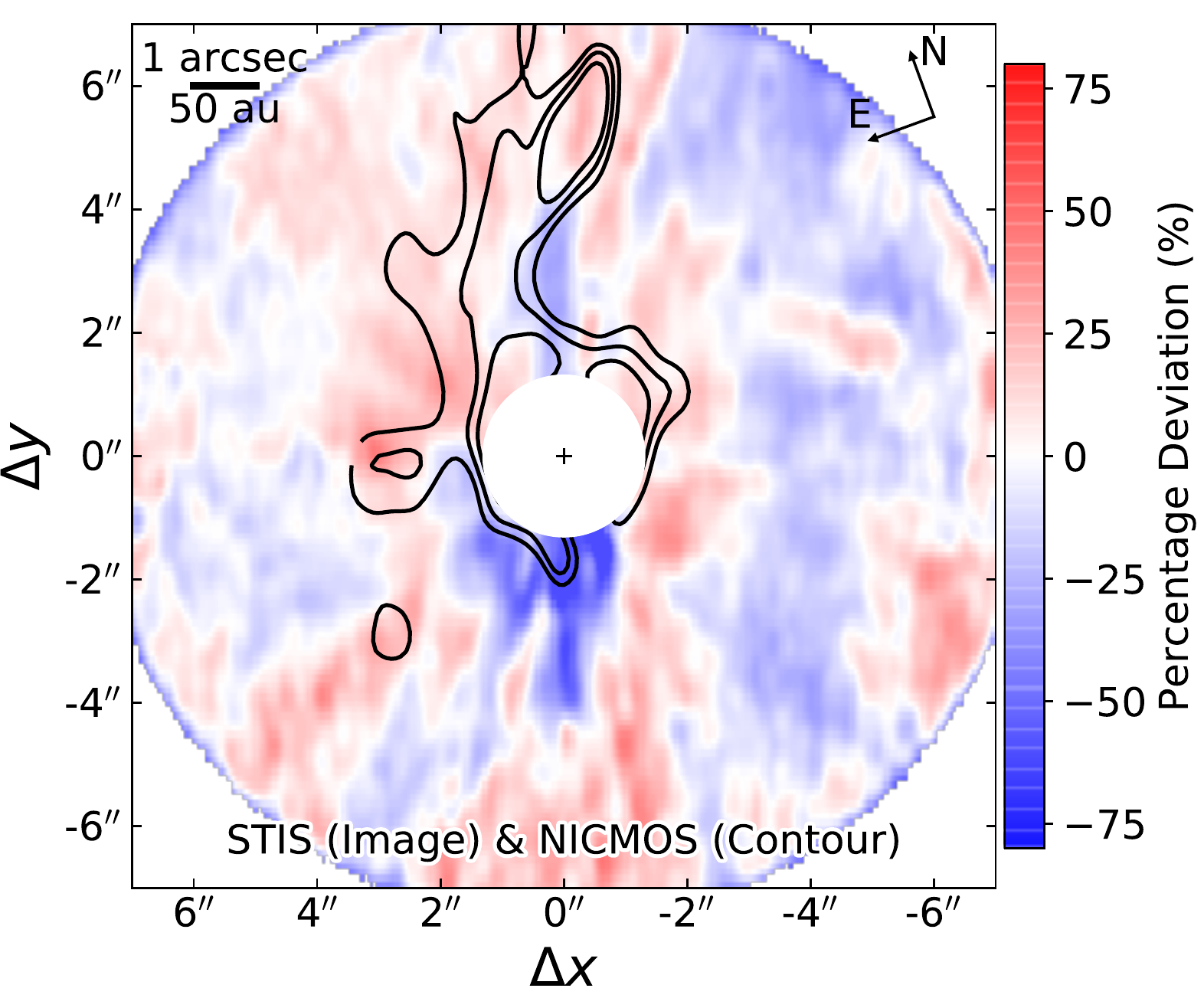}
\caption{Demonstration of the deviation of an identical SPF for the STIS halo.  {For comparison and illustration,} the NICMOS-NMF data are shown in  {black} contours {(arbitrary units)}. For the STIS halo, the NE and SW areas host overdensity regions, with the NE one likely matching that in the NICMOS-NMF data.}
\label{fig:spf-deviation}
\end{figure}

The 2-dimensional deviation from one SPF for both the STIS-cRDI and NICMOS-NMF are shown in Figure~\ref{fig:spf-deviation}. We observe overdensity regions in the NE and SW side of the STIS data at ${\sim}25\%$ level, with the STIS NE region {likely} matching the NICMOS NE overdensity region. {The under-density region to the NW region in the STIS image is likely influenced by the truncation of signal by STIS's Wedge B (Figure~\ref{fig:SNR}).}

Possible explanations for the deviation from {a uniform} SPF in the halo are as follows: (1) when the scattering properties of the dust are the same in the halo, the deviations are corresponding with local surface density variations; (2) when there is no density variation, then the dust's scattering properties are different; or (3) both effects are jointly affecting the SPFs in the halo.

{\it Stellar Encounter}: Based on the complicated structure of the halo, we investigate the scenario of whether the halo was created by stellar encounter events \citep[e.g.,][]{derosa19}. In the current epoch, the star at a separation of $11\farcs4$ to the southwest of HD~191089 in the STIS field of view (partially seen at the bottom right corner in Figure~\ref{fig:disk}a) is a background star. It is identified by {\it Gaia} DR2 with Source ID 6847146784384527872 at $d=1.06\pm0.06$ kpc \citep{gaia18}, thus it is not responsible for creating the halo. 

To trace the positions of the nearby stars in the past, we retrieve 44 stars that are within 5 pc from HD~191089 using the {\it Gaia} DR2 archive, and use the proper motions to linearly propagate the locations of the 44 stars in the past (${\sim}0.5$ Myr ago). We notice three stars that have the nearest projected approach from ${\sim}1.2$ pc to ${\sim}1.5$ pc, which happened between ${\sim}0.2$ Myr and ${\sim}0.35$ Myr from now\footnote{The results do not change using the $329$ stars within $10$~pc.}. For star encounter events, the closest approach are typically smaller than ${\sim}200$~au \citep[e.g.,][]{pfalzner03, pfalzner18}, and distinct features such as spiral arms dissipate beyond ${\sim}1000$ yr \citep{pfalzner03}. Therefore, if a star encounter event created the halo for the HD~191089 system, it should happen early in a cluster environment, where close encounters are more frequent \citep[e.g., the Solar System: ][]{pfalzner18}. Under this mechanism, the halo would have been dissipated.

\subsection{Dust Properties from Radiative Transfer Modeling}\label{sec-dis-mcrt-miedhs}

In this paper, based on previous efforts in debris disk modeling with DHS \citep[e.g.,][]{min10, milli17}, we adopted the DHS theory to model the dust in the HD~191089 ring for the GPI \Qr\ image in polarized light and the STIS image in total intensity. We cannot yet interpret the disk images with one model across different instruments. Although we have considered a limited combination of compositions, they span the range from refractory (carbon) to pure ice and even void (through porosity). The models consider a broad range of refractive index that encompass most standard astronomical compositions. Therefore, it is unlikely that our failure to find a good fit is solely due to not trying another composition.

In our modeling results, the STIS image favors Mie theory, and the GPI \Qr\ image favors DHS theory with maximum void fraction of ${\sim}21\%$. {The discrepancy in this parameter, which is the only additional parameter in DHS from Mie, indicates that the shape of the dust cannot be well approximated by either theory. In addition, although previous modeling attempts \citep[e.g.,][]{rodigas15, choquet17} have encountered that simple models cannot reproduce total intensity observation at multiple wavelengths, the analysis performed in this paper adds another dimension---polarization observation---to the complexity of disk modeling.

For some of the dust parameters (e.g., porosity, mass fraction of compositions), the discrepancies may be mathematically resolved using larger uncertainties by taking into account of correlated noise \citep[e.g.,][]{wolff17}. However, this resolution does not change the best-fit values, and in this way it would yield these parameters less informative since they can only take values in a limited range (i.e., from $0\%$ to $100\%$). The lack of meaning for these retrieved parameters further supports the fact that current models (i.e., Mie, DHS) cannot properly depict the debris disk images obtained at different wavelength and different observational techniques.}

The advances in dust descriptions may solve the discrepancies in the radiative transfer modeling of debris disks, including using laboratory measurements such as The Amsterdam-Granada Light Scattering Database \citep{munoz12}, or adopting advanced models for dust shapes and optical properties (e.g., discrete-dipole approximation: \citealp{purcell73, draine94, min06}; Rayleigh-Gans-Debye [RGD] theory: \citealp{sorensen01}; $T$-matrix method: \citealp{mishchenko96}; Gaussian random spheres: \citealp{muinonen96}; aggregation of small particles: \citealp{kempf99, tazaki16, tazaki18, arnold19}). {Given the more realistic descriptions of dust properties, the latter may help to resolve the discrepancies encountered with current dust models. For example, \citet{tazaki16} calculated the scattering phase functions for aggregates using the RGD theory, and found that backward scattering was underestimated in previous studies with simple models; \citet{arnold19} calculated blowout size for aggregates, and found that the size can vary as large as an order of magnitude for different particle models. Although these treatments may resolve the discrepancies, MCMC retrievals of dust properties for these advanced descriptions of dust are currently limited by computational power.}

For the HD~191089 system, the contribution from the halo in the STIS data is also calling for a more complex structural model. In addition, if the halo is not co-planar with the ring, then the halo will bias the SPF of the ring {and add another dimension of complexity to the problem}.

\section{Summary}\label{sec-conc}
In this paper, we report our detection and characterization of the HD~191089 debris disk by combining space- and ground-based instruments: {\it HST}/NICMOS, {\it HST}/STIS, and {\it Gemini}/GPI. Using these three instruments, we are able to study the disk in scattered light at three different wavelengths: $0.58\,\micron$ and $1.12\,\micron$ in total intensity, and $1.65\,\micron$ ($H$-band) in polarized intensity. In the scattered light images, we are able to identify two components in the debris disk system: a ring, and a fainter fan-like halo structure.  For the STIS and GPI \Qr\ images, we implement radiative transfer modeling to retrieve the dust information. Assuming the ring and the halo are coplanar, we summarize our findings as follows.

{\it Measurement}:
\begin{itemize}
	\item The HD~191089 system has two spatial components: one exo--Kuiper Belt ring from $26\pm4$~au to $78\pm14$~au, and an halo extending to $640\pm130$~au.  The center of ring does not have a significant offset from that of the star. The halo has an overall radial surface density power law index of $-0.68\pm0.04$, {with local variations indicative of the interaction with the interstellar medium}.
	\item The ring has an inclination of ${59^\circ}_{-2^\circ}^{+4^\circ}$, enabling the scattering phase function measurements for the two components from ${\sim}30^{\circ}$ to ${\sim}150^\circ$ in the STIS data. In the range of scattering angles probed by our observations, both forward and backward scattering are stronger for the dust in the halo than that in the ring. 
	\item {The polarization fraction curve calculated using $1.65\,\mu$m GPI \Qr\ and $1.12\,\mu$m NICMOS-NMF images does not have a clear trend, however it is possibly peaking at ${\sim}110^\circ$. Our result may be influenced by wavelength, and it can be better constrained with future total intensity observations in $H$-band.}
	\item  From the color of the dust derived from the NICMOS and STIS observations, the dust in the ring is redder than that in the halo. This is consistent with larger dust in the ring, which is also consistent with theoretical simulations that the ring serves as the ``birth ring'' for the smaller dust in the halo.
	\item In comparison with an identical SPF trend \citep{hughes18}, the SPFs of the dust in the HD~191089 system are likely deviating from the trend.
	\item If the ring is shaped by the strongest orbital resonances, the gravitational well is likely at $29.9\pm1.2$~au, resulting a $4:3$ resonance with the inner edge, and a $5:2$ resonance with the outer edge.
\end{itemize}

{\it Radiative Transfer Modeling}:\\
We use DHS theory to model the HD~191089 ring images observed with STIS and GPI. 
\begin{itemize}
	\item  Most of the extracted parameters are statistically not consistent with each other (e.g., composition, porosity). Specifically, the GPI \Qr\ ring favors DHS theory with ${\sim}21\%$ maximum void fraction, while the STIS ring favors Mie theory (i.e., DHS with $0\%$ void fraction). {This maximum void fraction parameter approximates the structure of the dust, which is the only parameter that is added between the two theories. The discrepancy for this parameter thus suggests that neither DHS nor Mie is a good approximation of the dust structure. However, both values are smaller than the best fit for interstellar dust \citep[0.7:][]{min07}, suggesting that dust properties are different in different enviroments. }
	\item {The discrepant dust parameters retrieved in our DHS radiative transfer modeling of the ring may be mathematically resolved with larger uncertainties. However, for the parameters that takes limited range of values from $0\%$ to $100\%$---e.g., porosity and mass fractions of compositions---large uncertainties will render these parameters less informative on dust properties. Advanced description of dust models such as aggregates are expected to physically resolve the discrepancies, however such MCMC analyses are currently limited by computational power.}
\end{itemize}

\acknowledgments
We appreciate the suggestions from the anonymous referee, which {significantly} improved this paper. B.R.~thanks the useful discussions with Xinyu Lu and the comments from Kevin Schlaufman which improved the paper. E.C.~acknowledges support from NASA through Hubble Fellowship grant HST-HF2-51355 awarded by STScI, operated by AURA, Inc.~under contract NAS5-26555, and support from HST-AR-12652, for research carried out at the Jet Propulsion Laboratory, California Institute of Technology. T.E.~was supported in part by NASA Grants NNX15AD95G/NEXSS, NNX15AC89G, and NSF AST-1518332. C.P.~acknowledges funding from the Australian Research Council via FT170100040 and DP180104235. G.D.~acknowledges support from NSF grants NNX15AD95G/NEXSS, AST-1413718 and AST-1616479. This research has made use of data reprocessed as part of the ALICE program, which was supported by NASA through grants HST-AR-12652 (PI: R. Soummer), HST-GO-11136 (PI: D.~Golimowski), HST-GO-13855 (PI: \'E.~Choquet), HST-GO-13331 (PI: L.~Pueyo), and STScI Director's Discretionary Research funds, and was conducted at STScI which is operated by AURA under NASA contract NAS5-26555. The input images to ALICE processing are from the recalibrated NICMOS data products produced by the Legacy Archive project, ``A Legacy Archive PSF Library And Circumstellar Environments (LAPLACE) Investigation,'' (HST-AR-11279, PI: G.~Schneider). Based on observations obtained at the Gemini Observatory, which is operated by the Association of Universities for Research in Astronomy, Inc., under a cooperative agreement with the NSF on behalf of the Gemini partnership: the National Science Foundation (United States), National Research Council (Canada), CONICYT (Chile), Ministerio de Ciencia, Tecnolog\'{i}a e Innovaci\'{o}n Productiva (Argentina), Minist\'{e}rio da Ci\^{e}ncia, Tecnologia e Inova\c{c}\~{a}o (Brazil), and Korea Astronomy and Space Science Institute (Republic of Korea).  This research project (or part of this research project) was conducted using computational resources (and/or scientific computing services) at the Maryland Advanced Research Computing Center (MARCC).
 
\facilities{HST (NICMOS, STIS), Gemini:South (Gemini Planet Imager)}

\software{{\tt corner.py} \citep{corner},  {\tt DebrisDiskFM} \citep{debrisdiskFM},  {\tt Debris Ring Analyzer} \citep{stark14}, {\tt emcee} \citep[Version 3.0rc1:][]{emcee}, {\tt MCFOST} \citep{pinte06, pinte09}, {\tt nmf\_imaging} \citep{nmfimaging}, {\tt pysynphot}  \citep{pysynphot}.}

\appendix

\section{The Scattering Phase Function}\label{appendix:spf}

\subsection{3D Cartesian Coordinates of the Disk System}\label{sec-appendix-3d}
To quantify the coordinate values, the $x$ and $y$ coordinates can be directly measured in the image of the system with $(x, y)\pm(\delta x, \delta y)$, where $\delta$ denotes the uncertainty of the parameters in this paper. In this section, given the other measured quantities of the system (i.e., inclination, and position angle of the semi-major axis), we obtain the $z$ coordinates for this system.

To determine the $z$ coordinates, we first set up the mathematical representation of the system. Let $O=(0, 0, 0)$ be the origin of the 3D Cartesian coordinate system, with $O$ placed at the geometric center of the debris disk, and unit vector $\hat{x}=(1, 0, 0)$ pointing towards West, $\hat{y} = (0, 1, 0)$ towards North, $\hat{z} = (0, 0, 1)$ pointing towards the observer. Let $\hat{n}_{\rm disk} = (a, b, c)$ denote the unit normal vector of the mid-plane of the debris disk system. Then, all the points on the disk mid-plane, which also contains the origin $O$, satisfies
\begin{equation}\label{eq-appendix-midplane-points}
ax + by + cz = 0,
\end{equation}
with $a^2 + b^2 + c^2 = 1$.\footnote{Note: the $a$ symbol in this section is not the dust size used in the main text.}

The inclination of the system, which is denoted by $\theta_{\rm inc}\pm\delta\theta_{\rm inc} \in [0^\circ, 90^\circ]$ and defined as the dihedral angle between the disk midplane and the $xOy$-plane, satisfies
\begin{equation*}
\cos \theta_{\rm inc} = \frac{\hat{n}_{\rm disk}\cdot\hat{n}_{xOy}}{\sqrt{||\hat{n}_{\rm disk}||^2||\hat{n}_{xOy}||^2}},
\end{equation*}
where $\cdot$ is the dot product between two vectors, and $\hat{n}_{xOy} = \hat{z}$ is the unit normal vector for the $xOy$-plane. The above equation becomes
\begin{equation}\label{eq-appendix-inc-c}
\cos \theta_{\rm inc} = \frac{(a, b, c) \cdot (0, 0, 1)}{\sqrt{(a^2+b^2+c^2)(0^2+0^2+1^2)}} = c.
\end{equation}

The position angle of the system, which is denoted by $\theta_{\rm PA}\pm\delta\theta_{\rm PA} \in [0^\circ, 180^\circ]$, is defined as the angle from North to the intersecting line between the system and the $xOy$-plane. For the intersecting line, it satisfies $ax + by = 0$ since the points on it are represented as $(x, y, 0)$. Let the mathematical slope angle of the line be $\theta_{\rm slope}$, which is defined as the counter-clockwise angle from $\hat{x}$ to the line. Then, we have the relationship between the mathematical slope angle and astronomical position angle,
\begin{equation*}
\theta_{\rm PA} = \theta_{\rm slope} - 90^\circ,
\end{equation*}
where
\begin{equation*}
\tan \theta_{\rm slope} = \frac{{\rm d}y}{{\rm d}x} = -\frac{a}{b},
\end{equation*}
therefore we have, 
\begin{equation}\label{eq-appendix-pa-ab}
\tan \theta_{\rm PA} = \tan (\theta_{\rm slope} - 90^\circ) = -\cot \theta_{\rm slope} = \frac{b}{a}.
\end{equation}

For the points on the disk mid-plane, we can substitute Equations~\eqref{eq-appendix-inc-c} and \eqref{eq-appendix-pa-ab} into Equation~\eqref{eq-appendix-midplane-points}, then we have the $z$-coordinate of the points as
\begin{align*}
z 	&= -\frac{1}{c}(ax + by) \\
	&= -\sec\theta_{\rm inc}  a\left(x +\frac{b}{a}y\right)\\
	&= \sec\theta_{\rm inc}\sin\theta_{\rm inc}\cos\theta_{\rm PA}(x+y\tan\theta_{\rm PA})\\
	&= \tan\theta_{\rm inc}(x\cos\theta_{\rm PA} + y\sin\theta_{\rm PA}) \numberthis \label{eq-appendix-z}.
\end{align*}
Assuming the measured parameters are independent, the squared uncertainty for $z$ is therefore
\begin{align*}
\delta^2 z 	= &\  \delta^2\left[{\tan\theta_{\rm inc}(x\cos\theta_{\rm PA} + y\sin\theta_{\rm PA})}\right]\\
		= &\  (x\cos\theta_{\rm PA} + y\sin\theta_{\rm PA})^2\delta^2(\tan\theta_{\rm inc})\\
		   & + \tan^2\theta_{\rm inc}\delta^2(x\cos\theta_{\rm PA} + y\sin\theta_{\rm PA})\\
		= &\  (x\cos\theta_{\rm PA} + y\sin\theta_{\rm PA})^2\sec^4\theta_{\rm inc}\delta^2(\theta_{\rm inc})\\
		&+\tan^2\theta_{\rm inc} \{[x^2\sin^2\theta_{\rm PA}\delta^2(\theta_{\rm PA})+\cos^2\theta_{\rm PA}\delta^2(x)]\\
		& +[y^2\cos^2\theta_{\rm PA}\delta^2(\theta_{\rm PA})+\sin^2\theta_{\rm PA}\delta^2(y)]\}\\
		=&\  \tan^2\theta_{\rm inc}\cos^2\theta_{\rm PA}\delta^2(x) + \tan^2\theta_{\rm inc}\sin^2\theta_{\rm PA}\delta^2(y)\\
		&+\sec^4\theta_{\rm inc}(x\cos\theta_{\rm PA} + y\sin\theta_{\rm PA})^2\delta^2(\theta_{\rm inc})\\
		&+ \tan^2\theta_{\rm inc}(x^2\sin^2\theta_{\rm PA}+y^2\cos^2\theta_{\rm PA})\delta^2(\theta_{\rm PA}) \numberthis \label{eq-appendix-z-unc}.
\end{align*}

Now combining the $(x, y)$ coordinates and the inclination and position angle of the system, we can obtain $z \pm \delta z$ from Equation~\eqref{eq-appendix-z} and the square-root of Equation~\eqref{eq-appendix-z-unc}. In this paper, for the HD~191089 disk, the $1\sigma$ uncertainties for $x$ and $y$ are estimated to be $0.33$ pixel\footnote{Scaled from a conservative $3\sigma$ uncertainty of 1 pixel}, and the uncertainties for $\theta_{\rm inc}$ and $\theta_{\rm PA}$ are obtained from the GPI \Qr\ image using the {\tt Debris Ring Analyzer} package by \citet{stark14}.

\subsection{Scattering Angle}\label{sec-appenidx-sa}
From the $(x, y, z)$ coordinates of the dust in the debris disk system, and given the position of the star at $(x_0, y_0, z_0)\pm(\delta x_0, \delta y_0, \delta z_0)$, we can then measure the scattering angle for the photons. A photon, which is emitted from the star and then interacts with the material at $(x, y, z)$, has an original direction of $\vec{r} = (x, y, z) - (x_0, y_0, z_0)$. The photon, when collected by the observer, has a final direction of $\hat{z}$. The scattering angle of this photon, which is defined and the angle between $\vec{r}$ and $\hat{z}$, is thus
\begin{align*}
\theta_{\rm scatter}  	&= \arccos\left(\frac{\vec{r}\cdot\hat{z}}{\sqrt{||\vec{r}||^2||\hat{z}||^2}} \right) \\
					&= \arccos\left[\frac{z-z_0}{\sqrt{(x-x_0)^2+(y-y_0)^2+(z-z_0)^2}}\right]\numberthis \label{eq-appendix-sa-form}.
\end{align*}
Substituting Equation~\eqref{eq-appendix-z} into the above equation, we have the scattering angle at the $(x, y)$ position in the detector frame (i.e., on the $xOy$-plane), 
\begin{widetext}
\begin{equation}
\theta_{\rm scatter} = \arccos\left\{\frac{\tan\theta_{\rm inc}(x\cos\theta_{\rm PA} + y\sin\theta_{\rm PA})-z_0}{\sqrt{(x-x_0)^2+(y-y_0)^2+\left[\tan\theta_{\rm inc}(x\cos\theta_{\rm PA} + y\sin\theta_{\rm PA})-z_0\right]^2}}\right\}. \label{eq-appendix-sa}
\end{equation}
\end{widetext}

Assuming the measured parameters are independent, the corresponding squared uncertainty for $\theta_{\rm scatter}$ is then
\begin{align*}
\delta^2\theta_{\rm scatter} = &\ \frac{1}{1-\left[\frac{z-z_0}{\sqrt{(x-x_0)^2+(y-y_0)^2+(z-z_0)^2}}\right]^2}\times\\
					&\delta^2\left[\frac{z-z_0}{\sqrt{(x-x_0)^2+(y-y_0)^2+(z-z_0)^2}}\right],
\end{align*}
denoting $v'\equiv v-v_0$ for $v\in\{x, y, z\}$ and thus $\delta^2 (v') = \delta^2(v)+\delta^2(v_0)$, then the above equation becomes
\begin{widetext}
\begin{align*}
\delta^2\theta_{\rm scatter} &= \frac{x'^2+y'^2+z'^2}{x'^2+y'^2}\delta^2\left(\frac{z'}{\sqrt{x'^2+y'^2+z'^2}}\right)\\
				&= \frac{x'^2+y'^2+z'^2}{x'^2+y'^2} \frac{z'^2}{x'^2+y'^2+z'^2}\left[\frac{\delta^2(z')}{z'^2} +\frac{\delta^2(\sqrt{x'^2+y'^2+z'^2})}{x'^2+y'^2+z'^2}\right]\\
				&=\frac{z'^2}{x'^2+y'^2}\left\{\frac{\delta^2(z')}{z'^2} +\frac{1}{x'^2+y'^2+z'^2}\left[\frac{1}{4(x'^2+y'^2+z'^2)}\delta^2(x'^2+y'^2+z'^2)\right]\right\}\\
				&= \frac{z'^2}{x'^2+y'^2}\left[\frac{\delta^2(z')}{z'^2} +\frac{x'^2\delta^2(x')+y'^2\delta^2(y')+z'^2\delta^2(z')}{(x'^2+y'^2+z'^2)^2}\right] \numberthis \label{eq-appendix-sa-unc-form}.
\end{align*}
\end{widetext}

Substituting the value and uncertainty of $z$ from Equations~\eqref{eq-appendix-z} and \eqref{eq-appendix-z-unc} into Equations~\eqref{eq-appendix-sa-form} and \eqref{eq-appendix-sa-unc-form}, then we can obtain the value and uncertainty for the sacttering angle, $\theta_{\rm scatter}\pm\delta\theta_{\rm scatter}$.

In this paper, the input uncertainties are obtained from the GPI \Qr\ image measured with the {\tt Debris Ring Analyzer} package \citep{stark14}. The SPF is then the scattering angle dependence of the flux density of the system at specific radial separations. The original measurements are then averaged to reduce measurement errors, and the final SPF is obtained by correcting the limb brightening effect \citep[by dividing the observed SPF by that of an isotropic disk model:][]{milli17}.

\section{Sampling from Posterior Distributions}\label{appendix:2smc}
\subsection{Glivenko-Cantelli Theorem}\label{sec:g-c}
In probability theory, for $n$ independent and identically distributed real-valued random variables ($X_1, X_2, \cdots, X_n\in\mathbb{R}$), the empirical cumulative distribution function (ECDF) is defined as\footnote{Note: the $x$ and $y$ symbols in \ref{appendix:2smc} are statistical variables, not Cartesian coordinates.} 
\begin{equation}
\hat{F}_n(x) = \frac{1}{n}\sum_{i=1}^n \mathbf{1}_{[X_i, \infty)}(x),
\end{equation}
where $ \mathbf{1}_{[X_i, \infty)}(x)$ is an indicator function which is equal to 1 only when $X_i \leq x < \infty$ (otherwise $\mathbf{1}_{[X_i, \infty)}(x) = 0$). For the ECDF, the Glivenko-Cantelli theorem \citep[e.g.,][]{chung2001book} describes its asymptotic relationship with the cumulative distribution function (CDF) of a random variable $X\in\mathbb{R}$, which is denoted by $F_X(x)$, that $\hat{F}_n(x)$ converges to $F_X(x)$ uniformly to $F_X(x)$ as $n\rightarrow\infty$ almost surely, i.e.,
\begin{equation}
||\hat{F}_n-F_X||_\infty \equiv \sup_{x\in\mathbb{R}}|\hat{F}_n(x)-F_X(x)|\xrightarrow{\rm a.s.}0. \label{eq:g-ct}
\end{equation}

\subsection{Probability Integral Transform (PIT)}\label{sec:pit}
In statistics, for a random variable $X\in\mathbb{R}$ with CDF $F_X(x)$, the PIT states that $Y = F_X(x)$ is uniformly distributed between 0 and 1 \citep{rosenblatt1952}, in the sense that
\begin{align*}
F_Y(y) &= \Pr(Y\leq y) \\
		&= \Pr[F_X(x) \leq y] \\
		&= \Pr[X\leq F_X^{-1}(y)] \\
		&= F_X[F_X^{-1}(y)]\\
		&= y, \numberthis \label{eq:pit}
\end{align*}
which is the CDF of a random variable that is uniformly distributed between 0 and 1, and $F_Y(y)$ is the CDF of the random variable $Y$. Based on this property, the PIT is used to sample distributions, especially the ones that do not have parametric expressions.

\subsection{Posterior as Prior}\label{sec:pap}
To use the marginal posterior distribution from the previous MCMC run as the prior for the next run, we transfer the information between the two MCMC runs by combining the Glivenko-Cantelli Theorem and the PIT:

First, convert discrete points to a continuous distribution: based on the Glivenko-Cantelli Theorem, for a specific random variable $X$ with a large number of samples, we can treat its marginal ECDF from the previous MCMC run as its CDF, then use the ECDF as the prior for the next MCMC run.

Second, sample from an ECDF: we first sample a standard uniform random quantile variable $Y\in[0, 1]$, then find the corresponding empirical quantile in the given ECDF, i.e., $\hat{F}_n^{-1}(Y)$. Using the PIT in Equation~\eqref{eq:pit}, we have 
\begin{equation}
\hat{F}_n^{-1}(Y) \sim \hat{F}_n.\label{eq:quantile-ecdf}
\end{equation}

Third, combining the Glivenko-Cantelli Theorem in Equation~\eqref{eq:g-ct} with the quantile distribution in Equation~\eqref{eq:quantile-ecdf}, we have
\begin{equation}
\hat{F}_n^{-1}(Y) \sim F_X,
\end{equation}
i.e., for a standard uniform random variable $Y$, its corresponding quantile for the ECDF of a random variable $X$ follows the distribution of $X$.

\subsection{GPI Posteriors as STIS Priors}\label{sec:pit-gpi-stis}
In this section, we establish the connection between the GPI \Qr\ image and the STIS total intensity image through radiative transfer modeling. We first obtain the posterior distribution of the disk parameters by radiative transfer modeling the GPI \Qr\ image with MCMC (Section~\ref{sec-mcmc-gpi}). We then calculate the marginal distributions from the MCMC posterior values\footnote{Note: the correlation of the parameters are ignored.}, and use the PIT to treat them as the priors when modeling the STIS image. 

\begin{figure*}[htb!]
\center
\includegraphics[width=\textwidth]{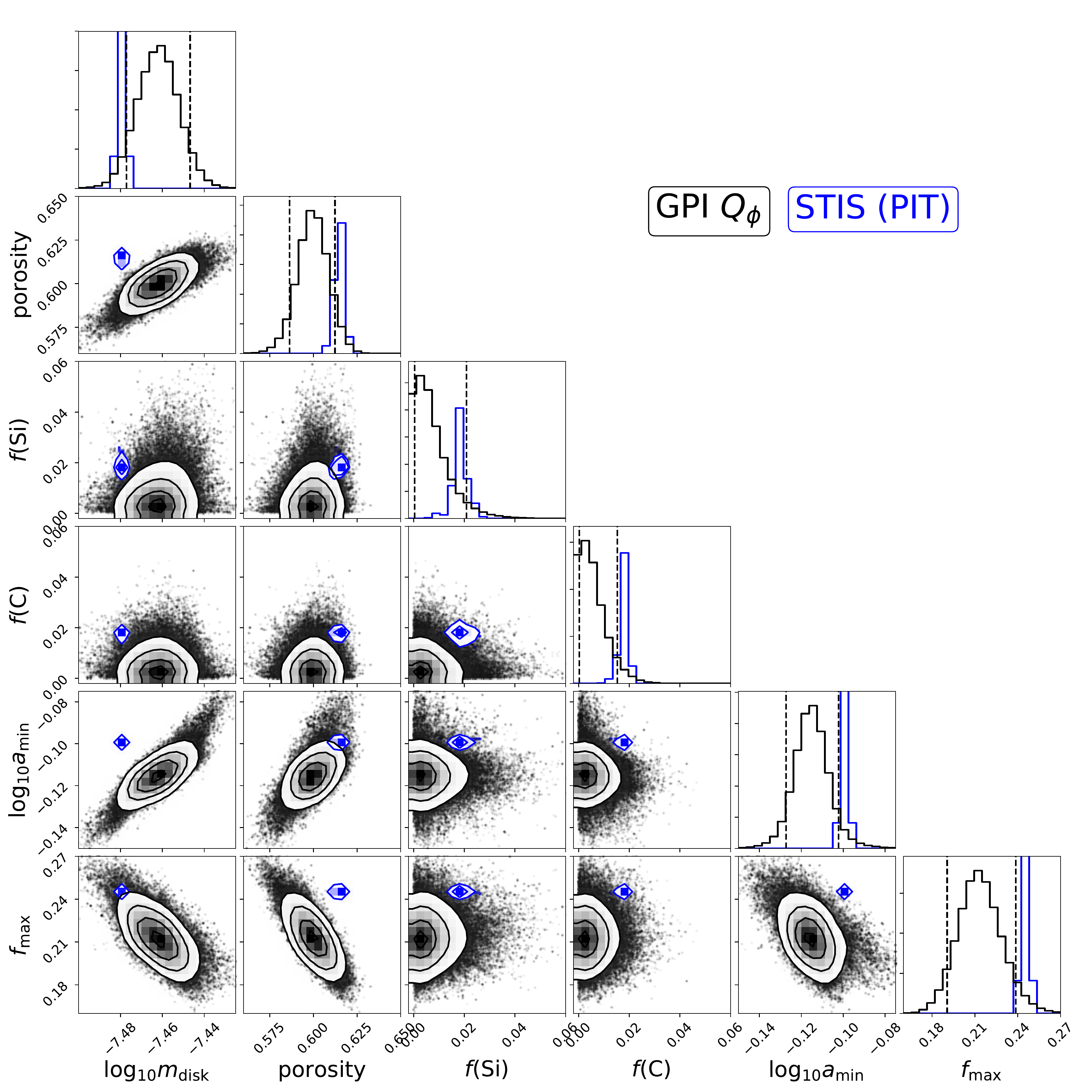}
\caption{Posterior distributions for STIS modeling with the PIT approach (blue). The marginal distribution of the GPI posteriors (black) are used as the priors for STIS fitting. The vertical dashed lines are the GPI 2.5th and 97.5th percentiles, which are the prior ranges. The majority of the STIS PIT posteriors lie around or beyond the 2.5th and 97.5th percentiles of the priors, indicating a trend of deviating from them.}
\label{fig:posterior-pit}
\end{figure*}

The posteriors with the PIT approach for the STIS image are presented in Figure~\ref{fig:posterior-pit}. In this section, our purpose is to demonstrate the statistical deviation of the posteriors from the GPI best-fit values.
We thus constrain the PIT sampling ranges to be between the 2.5th and 97.5th percentiles, which is analogous to the conventional definition of two-tailed statistical significance: $p$-value $>$ 0.05.

In this paper, we have ignored the correlated spatial noises in the images in our MCMC modeling; however, correlated noise is expected to increase the uncertainty of the extracted parameters \citep[e.g.,][]{czekala15, wolff17}. To better quantify the statistical deviation of the two sets of disk parameters that are extracted from the two disk images (Table~\ref{tab-mcfost-results}), rigorous treatment of the correlated noise is necessary \citep[e.g.,][]{wolff17}.

\newpage
\newpage
\bibliography{ms.bbl}
\end{CJK*}
\end{document}

%% file: AlphabetAuthors.txt
\author[0000-0001-5172-7902]{S. Mark Ammons}
\affiliation{Lawrence Livermore National Laboratory, 7000 East Avenue, Livermore, CA 94550, USA}

\author[0000-0003-4142-9842]{Megan Ansdell}
\affiliation{Astronomy Department, University of California, Berkeley, CA 94720, USA}

\author[0000-0001-6364-2834]{Pauline Arriaga}
\affiliation{Department of Physics \& Astronomy, 430 Portola Plaza, University of California, Los Angeles, CA 90095, USA}

\author[0000-0002-5407-2806]{Vanessa P. Bailey}
\affiliation{NASA Jet Propulsion Laboratory, California Institute of Technology, 4800 Oak Grove Drive, Pasadena, CA 91109, USA}

\author[0000-0002-7129-3002]{Travis Barman}
\affiliation{Lunar and Planetary Laboratory, The University of Arizona, Tucson, AZ 85721, USA}

\author[0000-0002-2731-0397]{Juan Sebasti{\'a}n Bruzzone}
\affiliation{Department of Physics and Astronomy, The University of Western Ontario, London, ON, N6A 3K7, Canada}

\author{Joanna Bulger}
\affiliation{Subaru Telescope, NAOJ, 650 North A{'o}hoku Place, Hilo, HI 96720, USA}

\author[0000-0001-6305-7272]{Jeffrey Chilcote}
\affiliation{Department of Physics, University of Notre Dame, 225 Nieuwland Science Hall, Notre Dame, IN 46556, USA}

\author[0000-0003-0156-3019]{Tara Cotten}
\affiliation{Department of Physics and Astronomy, University of Georgia, Athens, GA 30602, USA}

\author[0000-0002-4918-0247]{Robert J. De Rosa}
\affiliation{Astronomy Department, University of California, Berkeley, CA 94720, USA}

\author{Rene Doyon}
\affiliation{Institut de Recherche sur les Exoplan{\`e}tes, D{\'e}partement de Physique, Universit{\'e} de Montr{\'e}al, Montr{\'e}al, QC, H3C 3J7, Canada}

\author[0000-0002-0176-8973]{Michael P. Fitzgerald}
\affiliation{Department of Physics \& Astronomy, 430 Portola Plaza, University of California, Los Angeles, CA 90095, USA}

\author[0000-0002-7821-0695]{Katherine B. Follette}
\affiliation{Physics and Astronomy Department, Amherst College, 21 Merrill Science Drive, Amherst, MA 01002, USA}

\author[0000-0002-4144-5116]{Stephen J. Goodsell}
\affiliation{Gemini Observatory, 670 North A'ohoku Place, Hilo, HI 96720, USA}

\author[0000-0003-3978-9195]{Benjamin L. Gerard}
\affiliation{University of Victoria, Department of Physics and Astronomy, 3800 Finnerty Rd, Victoria, BC, V8P 5C2, Canada}
\affiliation{National Research Council of Canada Herzberg, 5071 West Saanich Rd, Victoria, BC, V9E 2E7, Canada}

\author{James R. Graham}
\affiliation{Astronomy Department, University of California, Berkeley, CA 94720, USA}

\author[0000-0002-7162-8036]{Alexandra Z. Greenbaum}
\affiliation{Department of Astronomy, University of Michigan, Ann Arbor, MI 48109, USA}

\author{J. Brendan Hagan}
\affiliation{Space Telescope Science Institute (STScI), 3700 San Martin Drive, Baltimore, MD 21218, USA}

\author[0000-0003-3726-5494]{Pascale Hibon}
\affiliation{Gemini Observatory, Casilla 603, La Serena, Chile}

\author[0000-0003-4653-6161]{Dean C.~Hines} 
\affiliation{Space Telescope Science Institute (STScI), 3700 San Martin Drive, Baltimore, MD 21218, USA}

\author[0000-0003-1498-6088]{Li-Wei Hung}
\affiliation{Natural Sounds and Night Skies Division, National Park Service, Fort Collins, CO 80525, USA}

\author[0000-0003-3715-8138]{Patrick Ingraham}
\affiliation{Large Synoptic Survey Telescope, 950N Cherry Avenue, Tucson, AZ 85719, USA}

\author{Paul Kalas}
\affiliation{Astronomy Department, University of California, Berkeley, CA 94720, USA}
\affiliation{SETI Institute, Carl Sagan Center, 189 Bernardo Avenue, Mountain View, CA 94043, USA}

\author[0000-0002-9936-6285]{Quinn Konopacky}
\affiliation{Center for Astrophysics and Space Science, University of California San Diego, La Jolla, CA 92093, USA}

\author{James E. Larkin}
\affiliation{Department of Physics \& Astronomy, 430 Portola Plaza, University of California, Los Angeles, CA 90095, USA}

\author[0000-0003-1212-7538]{Bruce Macintosh}
\affiliation{Kavli Institute for Particle Astrophysics and Cosmology, Department of Physics, Stanford University, Stanford, CA, 94305, USA}

\author{J\'er\^ome Maire}
\affiliation{Center for Astrophysics and Space Science, University of California San Diego, La Jolla, CA 92093, USA}

\author[0000-0001-7016-7277]{Franck Marchis}
\affiliation{SETI Institute, Carl Sagan Center, 189 Bernardo Avenue,  Mountain View, CA 94043, USA}

\author[0000-0002-4164-4182]{Christian Marois}
\affiliation{National Research Council of Canada Herzberg, 5071 West Saanich Rd, Victoria, BC, V9E 2E7, Canada}
\affiliation{University of Victoria, Department of Physics and Astronomy, 3800 Finnerty Rd, Victoria, BC, V8P 5C2, Canada}

\author[0000-0002-9133-3091]{Johan Mazoyer}
\altaffiliation{NASA Sagan Fellow}
\affiliation{NASA Jet Propulsion Laboratory, California Institute of Technology, 4800 Oak Grove Drive, Pasadena, CA 91109, USA}

\author[0000-0002-1637-7393]{Fran\c{c}ois M\'enard} 
\affiliation{Universit\'e Grenoble-Alpes, CNRS Institut de Plan\'etologie et d'Astrophysique (IPAG), F-38000 Grenoble, France}

\author[0000-0003-3050-8203]{Stanimir Metchev}
\affiliation{Department of Physics and Astronomy, The University of Western Ontario, London, ON, N6A 3K7, Canada}
\affiliation{Department of Physics and Astronomy, Stony Brook University, Stony Brook, NY 11794-3800, USA}

\author[0000-0001-6205-9233]{Maxwell A. Millar-Blanchaer}
\altaffiliation{NASA Hubble Fellow}
\affiliation{NASA Jet Propulsion Laboratory, California Institute of Technology, 4800 Oak Grove Drive, Pasadena, CA 91109, USA}

\author[0000-0002-8026-0018]{Tushar Mittal} 
\affiliation{Earth and Planetary Science Department, University of California, Berkeley, CA 94720, USA}

\author[0000-0003-2743-0943]{Margaret Moerchen}
\affiliation{American Geophysical Union, 2000 Florida Ave NW, Washington, DC 20009, USA}

\author[0000-0001-6975-9056]{Eric L. Nielsen}
\affiliation{Kavli Institute for Particle Astrophysics and Cosmology, Department of Physics, Stanford University, Stanford, CA, 94305, USA}

\author{Mamadou N'Diaye}
\affiliation{Universit\'e C\^ote d\'{}Azur, Observatoire de la C\^ote d\'{}Azur, CNRS, Laboratoire Lagrange, Bd de l\'{}Observatoire, CS 34229, 06304 Nice cedex 4, France}

\author[0000-0001-7130-7681]{Rebecca Oppenheimer}
\affiliation{Department of Astrophysics, American Museum of Natural History, New York, NY 10024, USA}

\author{David Palmer}
\affiliation{Lawrence Livermore National Laboratory, 7000 East Avenue, Livermore, CA 94550, USA}

\author{Jennifer Patience}
\affiliation{School of Earth and Space Exploration, Arizona State University, PO Box 871404, Tempe, AZ 85287, USA}

\author[0000-0001-5907-5179]{Christophe Pinte} 
\affiliation{Monash Centre for Astrophysics (MoCA) and School of Physics and Astronomy,
  Monash University, Clayton Vic 3800, Australia}
\affiliation{Universit\'e Grenoble-Alpes, CNRS Institut de Plan\'etologie et d'Astrophysique (IPAG), F-38000 Grenoble, France}

\author{Lisa Poyneer}
\affiliation{Lawrence Livermore National Laboratory, 7000 East Avenue, Livermore, CA 94550, USA}

\author[0000-0002-9246-5467]{Abhijith Rajan}
\affiliation{Space Telescope Science Institute (STScI), 3700 San Martin Drive, Baltimore, MD 21218, USA}

\author[0000-0003-0029-0258]{Julien Rameau}
\affiliation{Institut de Recherche sur les Exoplan{\`e}tes, D{\'e}partement de Physique, Universit{\'e} de Montr{\'e}al, Montr{\'e}al, QC, H3C 3J7, Canada}

\author[0000-0002-9667-2244]{Fredrik T. Rantakyr\"o}
\affiliation{Gemini Observatory, Casilla 603, La Serena, Chile}

\author[0000-0003-2233-4821]{Jean-Baptiste Ruffio}
\affiliation{Kavli Institute for Particle Astrophysics and Cosmology, Department of Physics, Stanford University, Stanford, CA, 94305, USA}

\author{Dominic Ryan}
\affiliation{Astronomy Department, University of California, Berkeley, CA 94720, USA}

\author[0000-0002-8711-7206]{Dmitry Savransky}
\affiliation{Sibley School of Mechanical and Aerospace Engineering, Cornell University, Ithaca, NY 14853, USA}

\author[0000-0002-6294-5937]{Adam C. Schneider}
\affiliation{School of Earth and Space Exploration, Arizona State University, PO Box 871404, Tempe, AZ 85287, USA}

\author[0000-0003-1251-4124]{Anand Sivaramakrishnan}
\affiliation{Space Telescope Science Institute (STScI), 3700 San Martin Drive, Baltimore, MD 21218, USA}

\author[0000-0002-5815-7372]{Inseok Song}
\affiliation{Department of Physics and Astronomy, University of Georgia, Athens, GA 30602, USA}

\author[0000-0003-2753-2819]{R\'emi Soummer} 
\affiliation{Space Telescope Science Institute (STScI), 3700 San Martin Drive, Baltimore, MD 21218, USA}

\author{Christopher Stark} 
\affiliation{Space Telescope Science Institute (STScI), 3700 San Martin Drive, Baltimore, MD 21218, USA}

\author[0000-0002-9121-3436]{Sandrine Thomas}
\affiliation{Large Synoptic Survey Telescope, 950N Cherry Avenue, Tucson, AZ 85719, USA}

\author[0000-0002-5902-7828]{Arthur Vigan}
\affiliation{Aix Marseille Univ, CNRS, CNES, LAM, Marseille, France}

\author{J. Kent Wallace}
\affiliation{NASA Jet Propulsion Laboratory, California Institute of Technology, 4800 Oak Grove Drive, Pasadena, CA 91109, USA}

\author[0000-0002-4479-8291]{Kimberly Ward-Duong}
\affiliation{School of Earth and Space Exploration, Arizona State University, PO Box 871404, Tempe, AZ 85287, USA}
\affiliation{Physics and Astronomy Department, Amherst College, 21 Merrill Science Drive, Amherst, MA 01002, USA}

\author[0000-0003-4483-5037]{Sloane Wiktorowicz}
\affiliation{Department of Astronomy, UC Santa Cruz, 1156 High St., Santa Cruz, CA 95064, USA }

\author[0000-0002-9977-8255]{Schuyler Wolff}
\affiliation{Leiden Observatory, Leiden University, P.O. Box 9513, 2300 RA Leiden, The Netherlands}

\author[0000-0001-7591-2731]{Marie Ygouf}
\affil{IPAC, California Institute of Technology, M/C 100-22, 770 S. Wilson Ave, Pasadena, CA 91125, USA}